\documentclass[%
 aip,
 amsmath,amssymb,
 reprint,%
]{revtex4-1}

\usepackage{graphicx}
\usepackage{dcolumn}
\usepackage{bm}
\usepackage{amsmath,amssymb,amsfonts,mathtools}
\usepackage{textcomp}
\usepackage{braket}
\usepackage{physics}
\usepackage{tabularx}
\usepackage{xcolor}
\usepackage{multirow}

\usepackage[utf8]{inputenc}
\usepackage[T1]{fontenc}
\usepackage{mathptmx}
\usepackage{etoolbox}

\usepackage{float}
\makeatletter
\let\newfloat\newfloat@ltx
\makeatother
\usepackage{algorithm}
\usepackage[noend]{algpseudocode}

\usepackage[T1]{fontenc}
\newcolumntype{Y}{>{\centering\arraybackslash}X}
\newcommand{\code}[1]{\texttt{#1}}

\def\BibTeX{{\rm B\kern-.05em{\sc i\kern-.025em b}\kern-.08em
    T\kern-.1667em\lower.7ex\hbox{E}\kern-.125emX}}

\makeatletter
\def\@email#1#2{
 \endgroup
 \patchcmd{\titleblock@produce}
  {\frontmatter@RRAPformat}
  {\frontmatter@RRAPformat{\produce@RRAP{*#1\href{mailto:#2}{#2}}}\frontmatter@RRAPformat}
  {}{}
}
\makeatother
\begin{document}

\preprint{AIP/123-QED}

\title{Thresholds for the distributed surface code in the presence of memory decoherence}

\author{Sébastian de Bone}
\affiliation{QuTech, Delft University of Technology, Lorentzweg 1, 2628 CJ Delft, The Netherlands}
\affiliation{QuSoft, CWI, Science Park 123, 1098 XG Amsterdam, The Netherlands}
\author{Paul Möller}
\affiliation{QuTech, Delft University of Technology, Lorentzweg 1, 2628 CJ Delft, The Netherlands}
\author{Conor E. Bradley}
\affiliation{QuTech, Delft University of Technology, Lorentzweg 1, 2628 CJ Delft, The Netherlands}
\affiliation{Pritzker School of Molecular Engineering, University of Chicago, Chicago, IL 60637, USA}
\author{Tim H. Taminiau}
\affiliation{QuTech, Delft University of Technology, Lorentzweg 1, 2628 CJ Delft, The Netherlands}
\author{David Elkouss}
\email[]{david.elkouss@oist.jp}
\affiliation{QuTech, Delft University of Technology, Lorentzweg 1, 2628 CJ Delft, The Netherlands}
\affiliation{Networked Quantum Devices Unit, Okinawa Institute of Science and Technology Graduate University, Okinawa, Japan}

\date{\today}

\begin{abstract}
In the search for scalable, fault-tolerant quantum computing, distributed quantum computers are promising candidates. These systems can be realized in large-scale quantum networks or condensed onto a single chip with closely situated nodes. We present a framework for numerical simulations of a memory channel using the distributed toric surface code, where each data qubit of the code is part of a separate node, and the error-detection performance depends on the quality of four-qubit Greenberger-Horne-Zeilinger (GHZ) states generated between the nodes. 
We quantitatively investigate the effect of memory decoherence and evaluate the advantage of GHZ creation protocols tailored to the level of decoherence. 
We do this by applying our framework for the particular case of color centers in diamond, employing models developed from experimental characterization of nitrogen-vacancy centers.
For diamond color centers, coherence times during entanglement generation are orders of magnitude lower than coherence times of idling qubits. These coherence times represent a limiting factor for applications, but previous surface code simulations did not treat them as such. Introducing limiting coherence times as a prominent noise factor makes it imperative to integrate realistic operation times into simulations and incorporate strategies for operation scheduling.
Our model predicts error probability thresholds for gate and measurement reduced by at least a factor of three compared to prior work with more idealized noise models. We also find a threshold of $4\cdot10^2$ in the ratio between the entanglement generation and the decoherence rates, setting a benchmark for experimental progress.  

\end{abstract}

\maketitle
\begin{figure}
\newlength{\xfigwd}
\setlength{\xfigwd}{\textwidth}
\end{figure}

\section{Introduction}\label{sec:introduction}
A \emph{distributed quantum computer}~\cite{groverQuantumTelecomputation1997, ciracDistributedQuantumComputation1999} realizes a large-scale processing system by using entanglement to link smaller quantum processing units. For example, the sub-units may be elements of a photonic chip, or form the nodes of a quantum network on a larger scale.~\cite{vanmeterPathScalableDistributed2016} Fault tolerance is naturally achieved by establishing the connectivity according to the architecture of a topological error-correction code. The distributed approach provides advantages in terms of scalability but is limited by the availability of high-quality entanglement. This is because entangled states are required for both inter-node operations and for detecting local errors with the error-correction code.~\cite{jiangDistributedQuantumComputation2007, vanmeterDistributedQuantumComputation2010, fujiiDistributedArchitectureScalable2012, liHighThresholdDistributed2012a, monroeLargeScaleModular2014, rametteFaulttolerantConnectionErrorcorrected2023}

\begin{figure*}
\centering
\includegraphics[width=\textwidth]{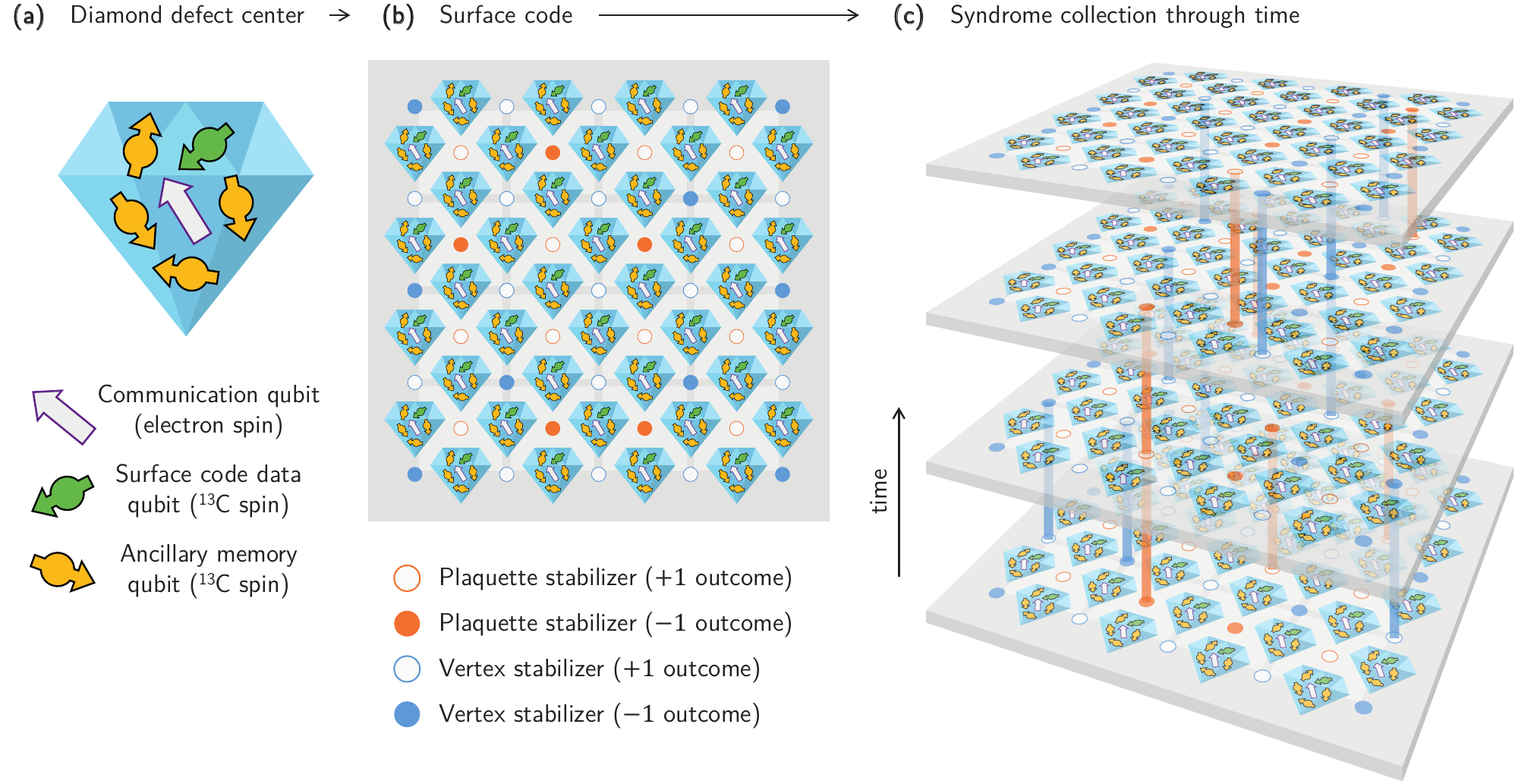}
\caption{\textbf{(a)} Schematic impression of a diamond defect center---also known as a diamond color center. The communication qubit is used to generate Bell pairs and to perform gates on the available qubits in the diamond color center. Out of all carbon spin memory qubits available, one is selected as the data qubit of the code. The other available memory qubits are used to store intermediate entangled states during the GHZ creation process. \textbf{(b)} Schematic impression of how a network of diamond color centers can be used to realize a surface code on a $4\times 4$ square lattice. Each center holds one data qubit of the error-correction code on one of its memory qubits. GHZ states are generated to measure the stabilizer operators of the code, resulting in an \emph{error syndrome} of $+1$ and $-1$ stabilizer measurement outcomes. \textbf{(c)} Stabilizer operators are measured consecutively in different time layers. A flip in stabilizer measurement outcome from one layer to the next is registered. The three-dimensional error syndrome that is created in this way is fed to an \emph{error syndrome decoder} to locate errors.}
\label{fig:connectivity_diamond_defect_centers}
\end{figure*}

In this paper, we focus on systems that are capable of generating remote two-qubit entanglement between pairs of connected nodes. There exist several physical systems suitable for generating this type of entanglement with optical interfaces.~\cite{awschalomQuantumTechnologiesOptically2018} Examples of this are ion traps, neutral atoms, and \emph{color centers} in the diamond lattice, such as nitrogen-vacancy (NV),~\cite{bernienHeraldedEntanglementSolidstate2013, Taminiau2014, riedelDeterministicEnhancementCoherent2017, 
Cramer2016, 
humphreysDeterministicDeliveryRemote2018, Bradley2019, Abobeih2019, rufResonantExcitationPurcell2021, pompiliRealizationMultinodeQuantum2021, abobeihFaulttolerantOperationLogical2022, bradleyRobustQuantumnetworkMemory2022, orphal-kobinOpticallyCoherentNitrogenVacancy2023} silicon-vacancy (SiV),~\cite{Sipahigil2016, sukachevSiliconVacancySpinQubit2017, nguyenIntegratedNanophotonicQuantum2019, nguyenQuantumNetworkNodes2019, knautEntanglementNanophotonicQuantum2023} or tin-vacancy (SnV)~\cite{iwasakiTinVacancyQuantumEmitters2017, rugarQuantumPhotonicInterface2021, debrouxQuantumControlTinVacancy2021} centers. As a concrete example, we investigate the distributed surface code with hardware-specific noise based on color centers, also known as \emph{defect centers}. We emphasize that the obtained insights are more general and that our simulation tools allow for implementing error models based on general hardware implementations. 

Diamond color centers host long-lived electron spins that exhibit coherent optical transitions, enabling their use as a \emph{communication} qubit. This qubit is used to create entanglement with other nodes and can address up to tens of proximal nuclear spins and or other electron spins occurring in the host material.~\cite{humphreysDeterministicDeliveryRemote2018, pompiliRealizationMultinodeQuantum2021, knautEntanglementNanophotonicQuantum2023} These nuclear spins can be used as local processing qubits---\textit{i.e.}, as a local memory to store and manipulate quantum states.~\cite{Bradley2019} Hereafter, in the context of diamond color centers, the term ``memory qubits'' specifically refers to such spins.

Physical systems suitable for distributed quantum computing can be operated as \emph{fault-tolerant} quantum computers by employing a subset of their memory qubits 
as data qubits of an \emph{error-correction code}, such as the toric surface code.~\cite{bravyiQuantumCodesLattice1998, dennisTopologicalQuantumMemory2002} The principle behind error-correction codes is that many physical qubits hold a smaller number of \emph{logical} states, and unwanted errors can be detected and corrected by measuring the \emph{stabilizer operators} of the code. For such a system, fault-tolerance goes hand-in-hand with the existence of \emph{thresholds} for local sources of error: if one manages to keep the error sources below their threshold values, one can make the logical error rate arbitrarily small by increasing the dimension of the error-correction code.

The toric code has a depolarizing \emph{phenomenological} error probability threshold of approximately 10\% to 11\%.\cite{brownedanLecturesTopologicalCodes2014} This error model assumes that all qubits of the code are part of the same quantum system, stabilizer measurements can be carried out perfectly, and the qubits experience depolarizing noise in between the stabilizer measurement rounds. A more precise analysis with a \emph{circuit-level} error model yields error probability thresholds between 0.9\% and 0.95\%.\cite{Nickerson2013} In this model, the stabilizer measurement circuit is simulated with noisy gates and measurements and it is implicitly assumed that the connectivity of the system allows direct two-qubit gates between adjacent qubits of the code topology. Therefore, this error model corresponds to a \emph{monolithic} architecture.

If one wants to implement the toric code in a network setting, where every data qubit of the code is part of a separate network node, the stabilizer operators can be measured with the aid of four-qubit Greenberger-Horne-Zeilinger (GHZ) states. These GHZ states can be created by \emph{fusing} three or more Bell pairs created between the involved nodes. 
Nickerson \textit{et al.}~analyzed the distributed toric code in this setting.~\cite{Nickerson2013,nickersonFreelyScalableQuantum2014} They included protocols with a relatively large number of \emph{entanglement distillation} steps that create high-quality GHZ states from imperfect Bell pairs. They found~\cite{Nickerson2013} thresholds for the local qubit operations between 0.6\% and 0.82\%---\textit{i.e.}, slightly below the monolithic thresholds. In their threshold calculations, Nickerson \textit{et al.}~do not explicitly consider circuit operation times and do not include qubit memory \emph{decoherence} during entanglement creation---\textit{i.e.}, the notion that the quality of the code's data qubits decays over time. However, in current physical systems of interest, decoherence during entanglement creation typically constitutes a large source of error.  
For state-of-the-art NV centers, coherence times during optical Bell pair generation are one to two orders of magnitude lower than estimated by Nickerson \textit{et al.}~\cite{reisererRobustQuantumNetworkMemory2016, Kalb2017a, kalbDephasingMechanismsDiamondbased2018} The influence of this decoherence is further increased by the reality that success probabilities per optical Bell pair generation attempt currently fall significantly short of unity.~\cite{hensenLoopholefreeBellInequality2015a, humphreysDeterministicDeliveryRemote2018}

Therefore, next to the errors in operations and in entangled states considered in Refs.~\onlinecite{Nickerson2013,nickersonFreelyScalableQuantum2014}, decoherence of quantum states over time emerges as the third primary source of noise for accurate assessment of distributed quantum computing systems. 
The influence of memory qubit decoherence during entanglement creation can be captured with the \emph{link efficiency} $\eta_\text{link}^*$.~\cite{bradleyRobustQuantumnetworkMemory2022} 
This parameter quantifies the average number of entangled pairs that can be generated within the coherence times. 

\begin{figure*}
\centering
\includegraphics[width=\textwidth]{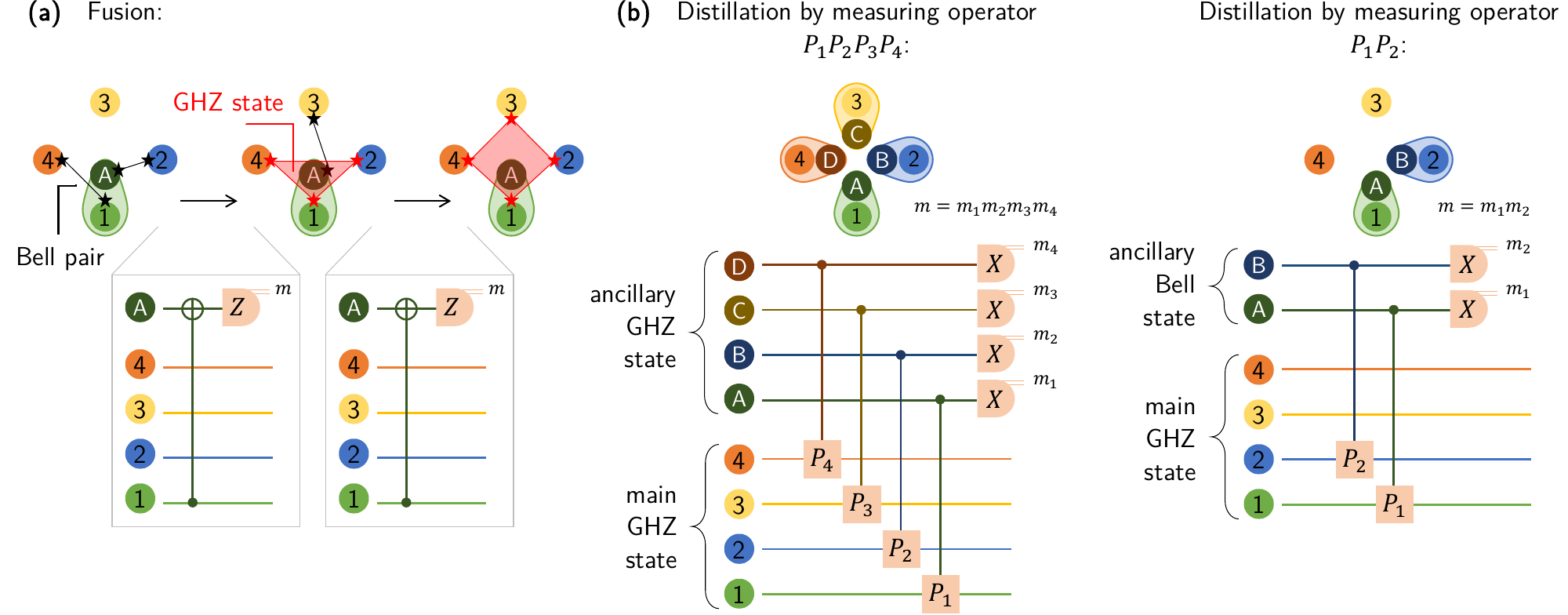}
\caption{\textbf{(a)} Example of the \emph{fusion} operation applied in the dynamic program of Alg.~\ref{alg:dynamic_program}. Fusion can be applied on any two states $\ket{\text{GHZ}_{n_1}}$ and $\ket{\text{GHZ}_{n_2}}$ that overlap in one or more network nodes. \textbf{(b)} Examples of the \emph{distillation} operation of Alg.~\ref{alg:dynamic_program}. Distillation consists on using one state of the form $\ket{\text{GHZ}_{n_1}}$ to measure a stabilizer of $\ket{\text{GHZ}_{n_2}}$---this could, \emph{e.g.}, be the operator $X_1 X_2 \dots X_{n_2}$, or $Z_1 Z_2$.}
\label{fig:fusion_and_distillation}
\end{figure*}

To investigate the influence of the coherence times, we develop a time-tracking simulator and implement realistic operation durations. 
Additionally, considering the pivotal role of the operation order in this new scenario, we formulate a strategy for scheduling operations. 
We find that, with realistic operation and coherence times, the thresholds with the GHZ generation protocols of Refs.~\onlinecite{Nickerson2013, nickersonFreelyScalableQuantum2014} disappear. We investigate the quantitative impact of memory decoherence and optimize over GHZ generation protocols with less distillation that can overcome this.
For a range of different coherence times during entanglement generation, we find two-qubit gate error and measurement error probability thresholds for diamond color centers up to $0.24$\%.
We find that fault-tolerance is reachable with $\eta_\text{link}^*\approx4\cdot10^2$. This improves on the prior results of $\eta_\text{link}^*=2\cdot10^{5}$ for the idealized time scale estimates of Nickerson \textit{et al.}~\cite{Nickerson2013} 
However, this link efficiency is still above the state-of-the-art hardware~\cite{bradleyRobustQuantumnetworkMemory2022}  reaching up to $\eta_\text{link}^*\approx 10$.

In the remainder of this paper, Sec.~\ref{sec:GHZ_creation_protocols} describes GHZ creation and distillation protocols necessary for the distributed surface code. Consequently, in Sec.~\ref{sec:distributed_toric_code}, we present the full cycle of stabilizer measurements of the surface code. In Sec.~\ref{sec:diamond_defect_center_model}, we describe error models that allow us to investigate a specific hardware implementation in the distributed surface code setting: diamond color centers. 
Finally, in Sec.~\ref{sec:results}, we investigate the parameter regimes necessary for fault tolerance with these error models.

\section{GHZ generation protocols}\label{sec:GHZ_creation_protocols}
As mentioned in Sec.~\ref{sec:introduction}, the stabilizer operators of a distributed quantum error-correcting code can be measured by consuming GHZ states. 
In the following, we discuss protocols that create GHZ states by combining Bell pairs. For each GHZ protocol, we identify two parameters. The first one is the minimum number of Bell pairs $k$ required to create the GHZ state. This number indicates the amount of distillation taking place in the protocol. 
The second one is the maximum number of qubits per node $q$ necessary to generate the GHZ state. 
We summarize prior work in Sec.~\ref{subsec:GHZ_protocols_in_literature}. In Sec.~\ref{subsec:dynamic_program_to_generate_GHZ_protocols}, we discuss our method for generating GHZ protocols.  

\subsection{Prior GHZ protocols}\label{subsec:GHZ_protocols_in_literature}
There is a plethora of prior work considering the generation and purification of GHZ states.~\cite{Murao98, maneva2002improved, Dur2003, aschauer2005multiparticle, glancyEntanglementPurificationAny2006, kruszynskaEntanglementPurificationProtocols2006, hostensHashingProtocolDistilling2006, hostensStabilizerStateBreeding2006, hoPurifyingGreenbergerHorneZeilingerStates2008, huberPurificationGenuineMultipartite2011, kuzmin2019scalable, riera2021entanglement, krastanov2021heterogeneous, rengaswamyEntanglementPurificationQuantum2024} Here, we focus on protocols that combine Bell pairs into a four-qubit GHZ state and discuss seven of them.

First, we consider two protocols that we used in an earlier study~\cite{bradleyRobustQuantumnetworkMemory2022}: the Plain ($k=3$, $q=2$) and Modicum ($k=4$, $q=2$) protocols. These protocols were designed to create a GHZ state with no distillation or only a single distillation step. The Plain protocol is the simplest protocol for creating a GHZ state from Bell pairs; it fuses three Bell pairs into a four-qubit GHZ state without any distillation. The Modicum protocol uses a fourth Bell pair to perform one round of distillation on the GHZ state.

On top of that, we consider five GHZ protocols found by Nickerson \textit{et al.}~in the context of distributed implementations of the toric code: Expedient ($k=22$, $q=3$) and Stringent ($k=42$, $q=3$) from Ref.~\onlinecite{Nickerson2013}, and Basic ($k=8$, $q=3$), Medium ($k=16$, $q=4$) and Refined ($k=40$, $q=5$) from Ref.~\onlinecite{nickersonFreelyScalableQuantum2014}.

\subsection{Dynamic program to generate GHZ protocols}\label{subsec:dynamic_program_to_generate_GHZ_protocols}
In this section, we present a method for optimizing GHZ creation with realistic noise models. We focus on creating GHZ states of the form $\ket{\text{GHZ}_n}=(\ket{0}^{\otimes n}+\ket{1}^{\otimes n})/\sqrt{2}$, where $n\in\{2,3,4\}$ represents the number of parties. We call $n$ the weight of the GHZ state. For convenience, we use the notation $\ket{\text{GHZ}_2}$ to describe a Bell state. 

We use a \emph{dynamic program} to optimize over the space of GHZ protocols. This program generates GHZ protocols by using two main operations: \emph{fusing} Bell pairs to create GHZ states, and \emph{distilling} or \emph{purifying} Bell or GHZ states by consuming other ancillary Bell pairs or GHZ states. Fig.~\ref{fig:fusion_and_distillation} depicts the two building blocks.

Distillation involves the use of an ancillary state to non-locally measure a stabilizer of the main Bell or GHZ state.~\cite{goyalPurificationLargeBicolorable2006} In this process, local control-Pauli gates between ancillary and main state qubits are followed by individual measurements of the ancillary state qubits in the Pauli-$X$ basis. Obtaining an even number of $-1$ measurement outcomes marks a successful distillation attempt. If distillation fails, the post-measurement state is discarded and (part of) the protocol has to be carried out again. 

Fusion is executed to create GHZ states out of Bell pairs and to create a larger GHZ state. A state of the form $\ket{\text{GHZ}_{n_1}}$ can be fused with a state $\ket{\text{GHZ}_{n_2}}$ by applying a CNOT gate between one qubit of both states and measuring out the target qubit in the Pauli-$Z$ basis. Obtaining a $+1$ measurement outcome results in the state $\ket{\text{GHZ}_{n_1+n_2-1}}$. A $-1$ measurement outcome leads to the same state after local Pauli-$X$ corrections. 

\begin{algorithm}[t]
\begin{algorithmic}
\Require {$n_\text{m}$: number of qubits of final GHZ state\\
		$k_\text{m}$: minimum number of Bell pairs used\\
		$V$: set with model parameters used\\
		$N_\text{b}$: protocols stored in buffer per step\\
		$N_\text{so}$: Monte Carlo shots used per protocol
		}
\For{$\{(n, k)\, |\, 2\leqslant n\leqslant n_\text{m}, \, n-n_\text{m}+k_\text{m}\geqslant k\geqslant n-1\}$}
	\State {\# Try all non-local measurement combinations.}
	\For{$g$ $\in$ \{stabilizers of $\ket{\text{GHZ}_n}$ \} }
		\State {$n'$ $\leftarrow$ weight of $g$}
		\For{$k' \in [n'-1, \, k-n+1]$}
			\For{$(p_1, p_2)\in [1, N_\text{b}]\times[1, N_\text{b}]$}
				\State {$\mathcal{P}_1$ $\leftarrow$ protocol $p_1$ in buffer at $(n,k-k')$}
				\State {$\mathcal{P}_2$ $\leftarrow$ protocol $p_2$ in buffer at $(n',k')$}
				\State {Construct binary tree protocol $\mathcal{P}_\text{new}$ that}
				\State {measures $g$ on $\mathcal{P}_1$ by consuming $\mathcal{P}_2$}
					\State {Construct protocol recipe $\mathcal{R}_\text{new}$ and evaluate}
					\State {quality over $N_\text{so}$ iterations times using $V$}
					\State {Store protocol if average performance}
					\State {is better than worst protocol in buffer}
			\EndFor
		\EndFor
	\EndFor
	\State {\# Try all fusion combinations.}
	\For{$n_2 \in [2, n-1]$}
		\State {$n_1\leftarrow n-n_2+1$}
		\For{$k_2 \in [n_2 -1, k-n+1]$}
			\State {$k_1\leftarrow k-k_2$}
			\For{$(p_1, p_2)\in [1, N_\text{b}]\times[1, N_\text{b}]$}
				\State {$\mathcal{P}_1$ $\leftarrow$ protocol $p_1$ in buffer at $(n_1,k_1)$}
				\State {$\mathcal{P}_2$ $\leftarrow$ protocol $p_2$ in buffer at $(n_2,k_2)$}
				\For{$(i, j)\in [1, n_1]\times [1, n_2]$}
					\State {Construct binary tree protocol $\mathcal{P}_\text{new}$ by}
					\State {fusing $\mathcal{P}_1$ at qubit $i$ and $\mathcal{P}_2$ at qubit $j$}
						\State {Construct protocol recipe $\mathcal{R}_\text{new}$ and eva-}
						\State {luate quality over $N_\text{so}$ iterations using $V$}
						\State {Store protocol if average performance}
						\State {is better than worst protocol in buffer}
				\EndFor
			\EndFor
		\EndFor
	\EndFor
\EndFor
\end{algorithmic}
\caption{Base dynamic program for GHZ protocols search.}
\label{alg:dynamic_program}
\end{algorithm}

\begin{figure*}
\centering
\includegraphics[width=\textwidth]{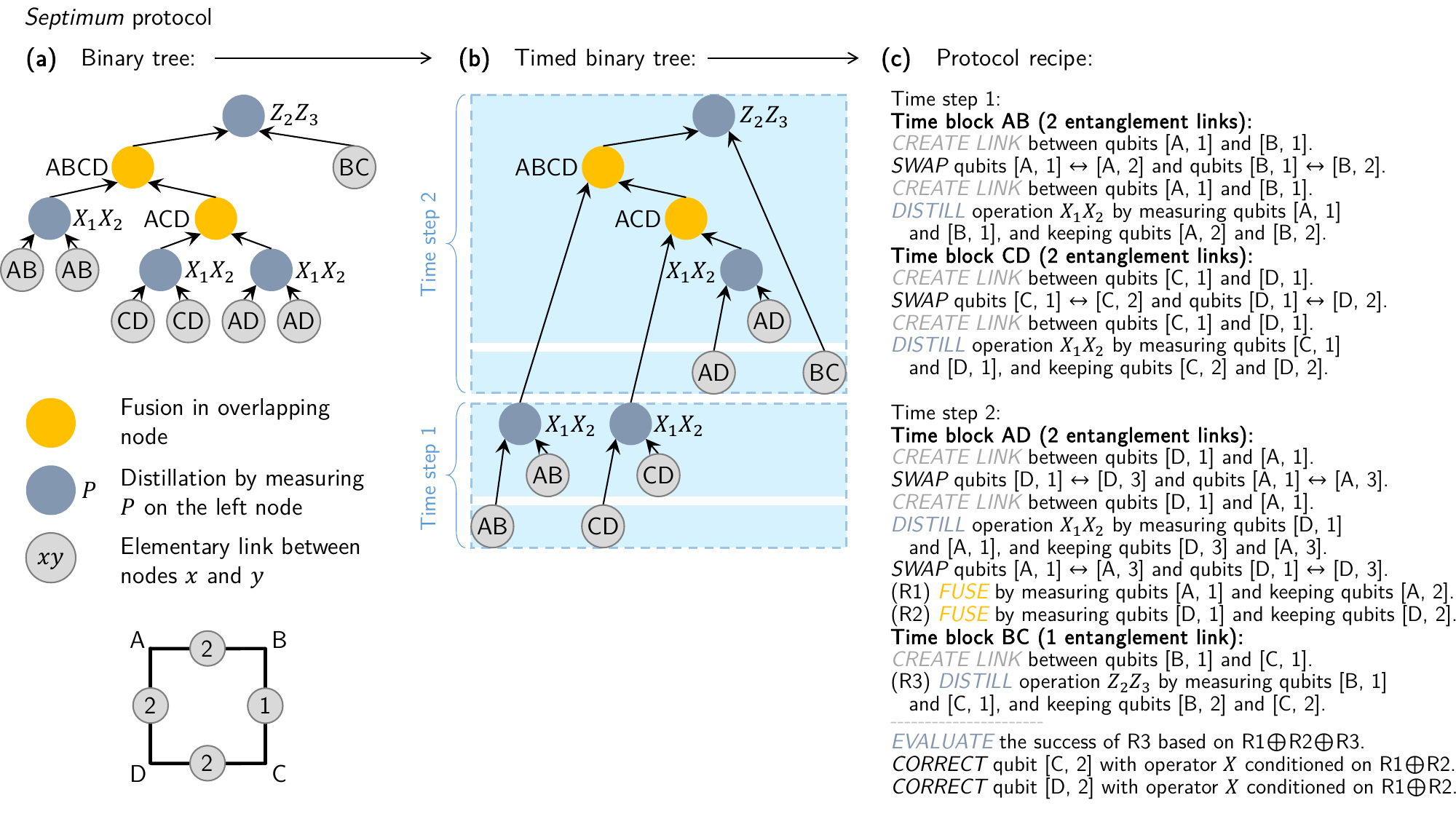}
\caption{\textbf{(a)} \emph{Binary tree} with $k=7$ found with the dynamic program of Alg.~\ref{alg:dynamic_program}: the \emph{Septimum} protocol. In this directed graph, the top vertex represents the final state. Each vertex describes how its corresponding state is created from a fusion or distillation operation involving its two children. At the origin of each branch, we find the elementary links: the Bell pairs. \textbf{(b)} We split the binary tree into multiple time steps that describe the order in which the protocol is carried out. The subtree involving the links between nodes $C$ and $D$ is identified as the part that we want to carry out first, since it is the left part of the largest branch of the binary tree. The subtree involving links between $A$ and $B$ is also added to time step 1, because it can be carried out in parallel. (See Sec.~\ref{subsec:dynamic_program_to_generate_GHZ_protocols} for more information.) \textbf{(c)} The timed binary tree is converted to an explicit set of operations: a \emph{protocol recipe}. Here, we also add necessary SWAP gates, conditional corrections for fusion operations, and evaluations of distillation operations. For distillation operations, we also add instructions in case of failure (not printed here). During the execution of this protocol, the system waits until all branches of a time step are completed before continuing to the next time step. This protocol recipe uses a maximum of $q=3$ qubits per network node to generate the GHZ state.}
\label{fig:protocol}
\end{figure*}

In Alg.~\ref{alg:dynamic_program}, we present a schematic, pseudo-code version of the dynamic program we used to generate and evaluate GHZ protocols. This algorithm is an expanded version of the dynamic program in Ref.~\onlinecite{deboneProtocolsCreatingDistilling2020}. In this algorithm, each protocol is created with either a fusion or a distillation operation that combines two smaller GHZ protocols encountered earlier in the search. The protocols created in this fashion can be depicted with a directed \emph{binary tree} graph. An example graph is given on the left side of Fig.~\ref{fig:protocol}. For the distillation steps in the binary tree diagrams, we consume the state on the right to distill the state on the left.

Each binary tree corresponds to multiple inequivalent protocols depending on the time ordering of the steps. We define a \emph{protocol recipe} as a set of instructions for implementing the protocol. The recipe includes the ordering of operations and state generation. An example of a protocol recipe can be seen on the right side of Fig.~\ref{fig:protocol}. This step was not required in previous research on distributed surface codes, as the noise models used in previous research did not include memory decoherence. 
Without a notion of time, the execution order of the tree's branches is irrelevant. 

As can be seen in Fig.~\ref{fig:protocol}, the conversion to a protocol recipe contains SWAP gates. These gates are required to match the connectivity constraints of our example hardware model---see Sec.~\ref{sec:diamond_defect_center_model} for more details. The SWAP gates should therefore not be considered as fundamental elements of these protocols and can be circumvented or neutralized in hardware systems with more operational freedom. We implement SWAP gates as three CNOT gates.

Whereas we did not optimize over the conversion from binary tree to protocol recipe, we considered two heuristics to limit the influence of decoherence and of SWAP gates. To limit decoherence, 
we prioritize creating larger branches of the tree. Here, a branch is defined as an element of the binary tree (\textit{i.e.}, an operation in the GHZ protocol) including all elements that (in)directly point towards it. The size of the branch is the number of elements it contains. Because, generally speaking, creating small branches is faster than creating large branches, this heuristic aims to minimize waiting times for completed intermediate branches of the GHZ protocol. 

The SWAP gate count can be limited by making sure a Bell pair that is consumed in a distillation step is the last state to be generated. This prevents the protocol from having to swap this state back and forth between the memory. For this reason, if two branches have equal size, we prioritize creating the left branch over the right one. 

In constructing the protocol recipe, we first use these heuristics to determine the order in which the elementary Bell pairs are generated---\textit{i.e.}, the leaves of the binary tree. By following this order, we then check for each Bell pair if other Bell pairs in non-overlapping network nodes can be generated simultaneously. Here, we prioritize based on proximity in the binary tree. We include instructions for distillation, fusion, and SWAP operations at the earliest possible point of execution. This approach gives rise to a unique conversion from binary tree to protocol recipe. More detailed descriptions of the protocol recipe construction and execution procedure can be found in Ref.~\onlinecite{deboneGHZDistillationProtocols2023} and the supplementary documents of the repository of Ref.~\onlinecite{deboneDataSoftwareUnderlying2024}.

While our dynamic program explores a large class of protocols, not all of the seven protocols that we introduced in the previous section can be generated. This is because, to suppress calculation time, the program limits distillation steps to operations that use an ancillary entangled state to non-locally measure a stabilizer operator of the main state, in a sequential manner. The protocols Refined, Expedient and Stringent, however, make use of the so-called \emph{double selection} distillation block~\cite{fujiiEntanglementPurificationDouble2009} that does not directly fit into this stabilizer distillation framework.~\cite{krastanovOptimizedEntanglementPurification2019, jansenEnumeratingAllBilocal2022, goodenoughNeartermDistillationProtocols2024}

\section{Distributed toric code}\label{sec:distributed_toric_code}
In this section, we discuss the steps of a distributed toric code and our approach for its simulation. 

\subsection{The toric surface code}\label{subsec:the_toric_surface_code}
In the toric surface code,~\cite{bravyiQuantumCodesLattice1998, dennisTopologicalQuantumMemory2002} data qubits are placed on the edges of an $L\times L$ lattice with periodic boundary conditions. It encodes two logical qubit states. The stabilizers of the code come in two forms: the product of $X$ operators on the four qubits surrounding every vertex of the lattice, and the product of $Z$ operators on the four qubits surrounding every face (or plaquette) of the code. 

\begin{figure*}
\centering
\includegraphics[width=\textwidth]{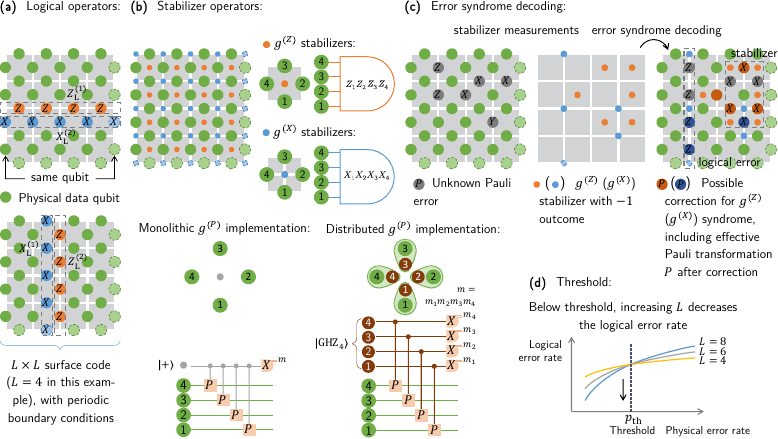}
\caption{General overview of the toric surface code. \textbf{(a)} The multi-qubit operators $Z_\text{L}^{(1,2)}$ and $X_\text{L}^{(1,2)}$ are the logical operators of the toric surface code. \textbf{(b)} The code's stabilizer generator operators are four-qubit operators surrounding every face ($g^{(Z)}$) and every vertex ($g^{(X)}$) of the lattice. In the distributed version, each stabilizer measurement requires the generation of an ancillary GHZ state between the nodes involved in the stabilizer measurement. \textbf{(c)} Unknown Pauli errors that appear on the physical data qubits can be tracked and corrected by measuring the stabilizer generators, and decoding the resulting \emph{error syndrome}. The unknown errors and the correction can still lead to a logical error. \textbf{(d)} The logical error probability increases with the physical error probability. This relation depends on the lattice sizes $L$. This can give rise to a physical error rate \emph{threshold}: below the threshold, the logical error rate decreases with the lattice size.}
\label{fig:surface_code_overview}
\end{figure*}

We consider a network topology with node connectivity matching the connectivity of the toric code lattice. We present a schematic impression in Fig.~\ref{fig:connectivity_diamond_defect_centers}. Each data qubit of the toric code is placed in a separate network node---\textit{e.g.}, in a separate diamond color center. The nodes have access to local memory qubits to create and store entangled links between them. Entangled links can be used to create four-qubit GHZ states, as described in Sec.~\ref{sec:GHZ_creation_protocols}, which are then consumed to measure the stabilizers of the code. Fig.~\ref{fig:surface_code_overview} shows a depiction of the procedure. 

The outcomes of the stabilizer measurements, known as the \emph{error syndrome}, are fed to a decoder to estimate the underlying error pattern. Here, we consider an implementation by Hu~\cite{huQuasilinearTimeDecoding2020, huOOPSurfaceCode} of the \emph{Union-Find}~\cite{delfosseAlmostlinearTimeDecoding2021} error decoder. 

We point out that we simulate the toric surface code as a logical quantum \emph{memory}, and do not consider its resilience against, \emph{e.g.}, logical operations or operations required for initializing logical information. This means we restrict the study to the code's ability to protect general logical states.

We opted for the toric surface code over the planar surface code (\textit{i.e.}, the surface code with boundaries) because, on top of the weight-4 stabilizer operators in the bulk, the planar code has lower-weight stabilizers at its boundaries. For distributed implementations, measuring these lower-weight stabilizers requires additional entanglement protocols. This makes simulating the planar code more complicated. Studies reveal that the introduction of boundaries typically has a limited effect on the code's threshold values, yet it is likely to result in a slower suppression of the logical error rates below these thresholds.~\cite{wangThresholdErrorRates2009}

\subsection{Distributed toric code simulation}\label{subsec:distributed_toric_code_calculations}
We split the simulation of the distributed toric code into two levels: simulation of the toric code's stabilizer measurements with the aid of GHZ states and simulation of $L$ rounds of stabilizer measurements of the code itself. 

The first level characterizes the stabilizer measurement. 
To this aim, we use Monte Carlo simulation to construct the (average) \emph{Choi state} associated with using the protocol's GHZ state to measure the plaquette or star stabilizer operator on half of the maximally entangled state. Exploiting channel-state duality, the Choi state from each iteration is converted to a \emph{superoperator} describing the stabilizer measurement channel. A formal introduction on channel characterization with a maximally entangled state can be found in Ref.~\onlinecite{wilde2013quantum}. The superoperator construction is described in detail in App.~\ref{app:superoperator_calculations}. 

\begin{figure*}
\centering
\includegraphics[width=\textwidth]{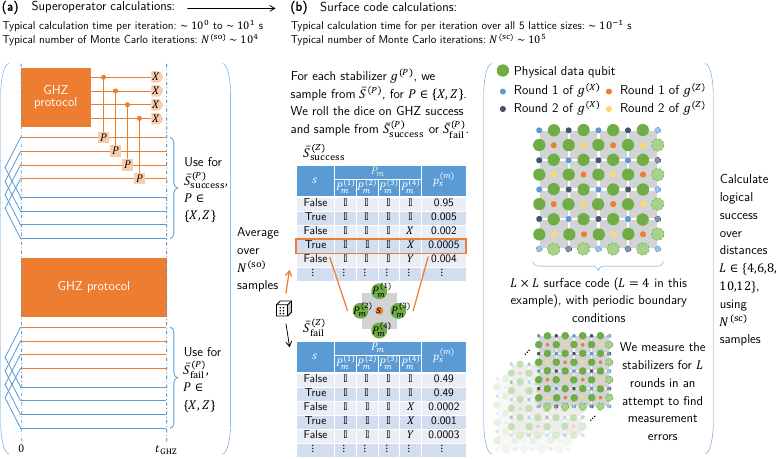}
\caption{Calculation process for error probability threshold simulations of the distributed surface code. This process is called for each specific protocol recipe and parameter set combination. The full calculation consists of two levels of Monte Carlo simulations: \textbf{(a)} the calculation of the superoperators $\overline{S}_\text{success}^{(P)}$ and $\overline{S}_\text{fail}^{(P)}$, for $P\in\{X,Z\}$, and \textbf{(b)} the surface code simulations using these superoperators.}
\label{fig:calculation_process}
\end{figure*}

The second level is a toric code simulator that takes as noise model the average superoperator obtained in the first level. 
Following previous research,~\cite{Nickerson2013} we consider a stabilizer measurement cycle consisting of two rounds of plaquette stabilizer measurements and two rounds of star stabilizer measurements. This is because the constraint that each network node only has one single communication qubit in our example hardware model makes it impossible to simultaneously generate entanglement for overlapping stabilizers---see Sec.~\ref{sec:diamond_defect_center_model} for more details. By splitting the full cycle up into four rounds, each defect center becomes part of exactly one stabilizer measurement per round.
This process is schematically depicted in Fig.~\ref{fig:calculation_process}. We note that a different hardware model could require, or benefit from, a different scheduling of the stabilizer measurements as the one used here. 

Due to entanglement distillation steps in the GHZ creation protocol, GHZ generation is probabilistic. To fix the duration of the rounds we impose a ``GHZ cycle time'' $t_\mathrm{GHZ}$: if GHZ generation is not successful within this time, it aborts. In that case, the corresponding stabilizer operator can not be measured. This information could be given to the decoder in the form of an erasure symbol. However, to leverage existing decoders, we opt to duplicate the last measured value of the stabilizer. This choice is suboptimal and better thresholds could be expected for decoders that can handle erasures and noisy measurements. The GHZ cycle time is a free parameter that we explore in Sec.~\ref{sec:results}. In App.~\ref{subsec:ghz_cycle_time_settings}, we describe a heuristic-driven approach for selecting a suitable GHZ cycle time.

To model GHZ generation failures, at the first level, we construct two separate average superoperators per stabilizer type: a \emph{successful} superoperator $\overline{S}_\text{success}$ for iterations where the GHZ state is created within $t_\mathrm{GHZ}$, and an \emph{unsuccessful} superoperator $\overline{S}_\text{fail}$ for iterations where the GHZ could not be created. Both superoperators incorporate the influence of decoherence up to the cycle time on the code's data qubits.

\section{Diamond color center model}\label{sec:diamond_defect_center_model}
In this section, we present an overview of the noise models and model parameters used in combination with the distributed surface code structure of Sec.~\ref{subsec:the_toric_surface_code}. The noise models are based on the nitrogen-vacancy (NV) center.~\cite{Bradley2019, pompiliRealizationMultinodeQuantum2021, bradleyRobustQuantumnetworkMemory2022} More information about the experimental characterization can be found in App.~\ref{sec:experimental_characterization}. More details about the models can be found in App.~\ref{sec:noise_models}. The parameter values can be found in Table~\ref{table:simulation_parameters}. 

In our model, qubits undergo both \emph{generalized amplitude damping} and \emph{phase damping} noise,~\cite{Nielsen2000} with separate $T_1$ and $T_2$ times. The generalized amplitude damping channel decoheres to the maximally mixed state $(\ketbra{0}{0}+\ketbra{1}{1})/2$. Decoherence of the electron qubit is governed by $T1_\text{idle}^\text{e}$ and $T2_\text{idle}^\text{e}$ coherence times. For the memory qubits, we use different coherence times $T1_\text{idle}^\text{n}$ and $T2_\text{idle}^\text{n}$ for when the node is idling, versus $T1_\text{link}^\text{n}$ and $T2_\text{link}^\text{n}$ during optical entanglement creation. This is because the required operations on the electron qubit typically induce additional decoherence on the memory qubits.~\cite{pompiliRealizationMultinodeQuantum2021, bradleyRobustQuantumnetworkMemory2022} 
The Kraus operators of both channels can be found in App.~\ref{subsec:decoherence}.

To mitigate decoherence from quasi-static noise processes in diamond color center experiments, \emph{dynamical decoupling} is typically interleaved into gate sequences on both the electron and nuclear spins.~\cite{Bradley2019} 
Here, the coherence times that we consider are also those achieved for NV center spin registers with dynamical decoupling.~\cite{Bradley2019} Consequently, in our numerical models, gate operations must be performed only between two consecutive dynamical decoupling pulses---\textit{i.e.}, at the refocusing point of the qubit spins involved in the operations. We define the center-to-center time of consecutive refocusing points as $t_\mathrm{DD} =  t_\mathrm{pulse} + 2n_\mathrm{DD}t_\mathrm{link}$, where $t_\mathrm{pulse}$ is the time of a $\pi$-pulse, $t_\mathrm{link}$ is the duration of a single Bell pair generation attempt, and $n_\mathrm{DD}$ is a fixed number of Bell pair generation attempts. In App.~\ref{subsec:dynamical_decoupling_sequence_length}, we discuss how $n_\mathrm{DD}$ is optimized. In our model, we assume that all memory qubits of a node are decoupled synchronously. 

We assume that each diamond color center only possesses a single communication qubit. Within each node, we further assume that measurements are only possible on its communication qubit, and local (\textit{i.e.}, intra-node) two-qubit gates always require the communication qubit to be the control qubit. These requirements mean we have to use SWAP gates to measure the state of a memory qubit or to use a memory qubit as the control of a two-qubit gate, as can be seen in Fig.~\ref{fig:protocol}. Lastly, we assume that a Bell pair between two centers can only be generated between the communication qubits of the two centers. We note that, in general, one could design (combinations of) diamond color center nodes with multiple communication qubits. Whereas this limits the number of SWAP gates required for distillation, this gives rise to extra requirements on the memory robustness of the communication qubits. 

The generation of Bell pairs between two color centers is modeled as a probabilistic process with a success probability $p_\text{link}$ and time $t_\text{link}$ per attempt. We constructed an analytic model to calculate $p_\text{link}$ and the Bell pair density matrix after success, both for the \emph{single-click} (\textit{i.e.}, the \emph{single-photon})~\cite{cabrilloCreationEntangledStates1999} entanglement protocol and for the \emph{double-click} (\textit{i.e.}, the \emph{two-photon})~\cite{barrettEfficientHighfidelityQuantum2005a} entanglement protocol. The following five noise sources are included in this analytic model: the preparation error of the initial spin-photon state $F_\text{prep}$, the probability of an excitation error $p_\text{EE}$, a parameter $\lambda$ based on the standard deviation of the phase uncertainty due to the path-length difference between the two arms (all modeled as dephasing channels on the involved qubits), the photon indistinguishability $\mu$ for each Bell state measurement (\emph{i.e.}, Hong-Ou-Mandel visibility, modeled with altered measurement operators~\cite{dahlbergLinkLayerProtocol2019}), and the total photon detection probability $\eta_\mathrm{ph}$ in each arm (modeled with an amplitude damping channel). 

For the double-click protocol, we assume the phase uncertainty to not be relevant and set $\lambda=1$. The parameters $F_\text{prep}$, $p_\text{EE}$ and $\mu$ affect the fidelity $F_\text{link}$ of the Bell pair state of the double-click protocol, whereas the parameter $\eta_\mathrm{ph}$ limits the success probability $p_\text{link}$ of a single entanglement attempt. For the single-click protocol, the fidelity $F_\text{link}$ is additionally influenced by $\eta_\mathrm{ph}$ and $\lambda$, and $p_\text{link}$ depends on the indistinguishability $\mu$. A full description of the density matrices and success probabilities of both entanglement protocols can be found in App.~\ref{subsec:bell_state}. 

The link efficiency $\eta_\text{link}^*$, introduced in Sec.~\ref{sec:introduction}, is defined in terms of the parameters $p_\text{link}$, $t_\text{link}$, $T1_\text{link}^\text{n}$ and $T2_\text{link}^\text{n}$ as
\begin{equation}\label{eq:def_link_efficiency}
\eta_\text{link}^*=\frac{2p_\text{link}}{t_\text{link}((T1_\text{link}^\text{n})^{-1}+(T2_\text{link}^\text{n})^{-1})}.
\end{equation}

We assume that all operations take a finite amount of time. The time durations can be found in Table~\ref{table:simulation_parameters}. We neglect the influence of classical communication times, as we consider distances between network nodes to be relatively small, but include synchronization of network nodes when classical communication is required. Furthermore, we assume single-qubit gates to be noiseless, while noise in two-qubit gates is modeled with a depolarizing channel with probability $p_\text{g}$ (see App.~\ref{subsec:gate_and_measurement_noise}). We model imperfect measurements by flipping the measurement outcome with probability $p_\text{m}$.

\section{Results}\label{sec:results}
In this section, we investigate the sensitivity of the distributed toric surface code performance with respect to several physical parameters of the diamond color center hardware model. In particular, we investigate the influence of two-qubit gate and measurement noise, the entanglement success probability, the coherence times during entanglement generation, and the quality of the generated Bell pairs on the noise thresholds. 
The threshold values $p_\textrm{th}$ are determined with fits of the logical success rates versus the lattice size ($L$) and local two-qubit gate and measurement error rate ($p_\text{g}=p_\text{m}$). The details of the fitting procedure can be found in App.~\ref{app:fitting_procedure_threshold_plots}. 

\begin{table}
\centering
\begin{tabularx}{\columnwidth}{|Y||Y|Y|Y|Y|}
\multicolumn{1}{c}{}& \multicolumn{1}{c}{\textbf{State-of-}}        & \multicolumn{1}{c}{\textbf{Fig.~\ref{fig:bell_state_quality_plot}}}        & \multicolumn{1}{c}{\textbf{Fig.~\ref{fig:link_efficiency_plot}}}        & \multicolumn{1}{c}{\textbf{Fig.~\ref{fig:bell_succ_prob_plot}}} \\ 
\multicolumn{1}{c}{}& \multicolumn{1}{c}{\textbf{the-art~\cite{Bradley2019, bradleyRobustQuantumnetworkMemory2022}}}        & \multicolumn{1}{c}{\textbf{}}        & \multicolumn{1}{c}{\textbf{}}        & \multicolumn{1}{c}{\textbf{}} \\ 
\multicolumn{5}{c}{\textit{Bell pair model input}}         \\ [0.5ex]
\hline
Protocol           & Single-click~\cite{cabrilloCreationEntangledStates1999} & \multicolumn{3}{c|}{Double-click~\cite{barrettEfficientHighfidelityQuantum2005a}} 	\\
\cline{2-5}
$F_\text{prep}$   & $0.99$\;\cite{hensenLoopholefreeBellInequality2015a, rozpedekNeartermQuantumrepeaterExperiments2019}         & \multicolumn{3}{c|}{$0.999$      }               \\
\cline{2-5}
$p_\text{EE}$      & $0.04$\;\cite{humphreysDeterministicDeliveryRemote2018}         & $p_\text{EE}(f_\phi)$         & \multicolumn{2}{c|}{$0.01$}              \\
\cline{2-5}
$\mu$              & $0.9$\;\cite{humphreysDeterministicDeliveryRemote2018, pompiliRealizationMultinodeQuantum2021}          & \multicolumn{3}{c|}{$0.95$}              \\
\cline{2-5}
$\lambda$          & $0.984$\;\cite{humphreysDeterministicDeliveryRemote2018, dahlbergLinkLayerProtocol2019}           & \multicolumn{3}{c|}{$1$}              \\
\cline{2-5}
$\eta_\mathrm{ph}$             & $0.0046$\;\cite{coopmansNetSquidNETworkSimulator2021}      & \multicolumn{2}{c|}{$0.4472$}             & $\eta_\mathrm{ph}(f_\eta)$             \\
\cline{2-5}
\hline
\multicolumn{5}{c}{} \\ [-5pt]
\multicolumn{5}{c}{\textit{Bell pair model output}}                                    \\ [0.5ex]
\hline
$p_\text{link}$    & $0.0001$       & \multicolumn{2}{c|}{$0.1$}            & $p_\text{link}(f_\eta)$              \\
\cline{2-5}
$F_\text{link}$    & $0.8966$           & $F_\text{link}(f_\phi)$       & \multicolumn{2}{c|}{$0.9526$}            \\
\hline
\multicolumn{5}{c}{} \\ [-5pt]
\multicolumn{5}{c}{\textit{Operation durations}}                                     \\ [0.5ex]
\hline
$t_\text{link}$    & \multicolumn{4}{c|}{$6\cdot10^{-6}$ s}              \\
\cline{2-5}
$t_\text{meas}$   & \multicolumn{4}{c|}{$4\cdot10^{-6}$ s}              \\
\cline{2-5}
$t_{X,Y}^\text{e}$ & \multicolumn{4}{c|}{$0.14\cdot10^{-6}$ s}             \\
\cline{2-5}
$t_{X,Y}^\text{n}$ & \multicolumn{4}{c|}{$1.0\cdot10^{-3}$ s}                  \\
\cline{2-5}
$t_{Z,H}^\text{e}$ & \multicolumn{4}{c|}{$0.1\cdot10^{-6}$ s}             \\
\cline{2-5}
$t_{Z,H}^\text{n}$ & \multicolumn{4}{c|}{$0.5\cdot10^{-3}$ s}                   \\
\cline{2-5}
$t_{\text{C}Z,\text{C}X,\text{C}iY}$ & \multicolumn{4}{c|}{$0.5\cdot10^{-3}$ s}                    \\
\cline{2-5}
$t_\text{SWAP}$ & \multicolumn{4}{c|}{$1.5\cdot10^{-3}$ s}             \\
\hline
\multicolumn{5}{c}{} \\ [-5pt]
\multicolumn{5}{c}{\textit{Decoherence}}                                     \\ [0.5ex]
\hline
$T1_\text{idle}^\text{n}$ & \multicolumn{4}{c|}{$300$ s}             \\
\cline{2-5}
$T1_\text{link}^\text{n}$ & $0.03$ s\;\cite{pompiliRealizationMultinodeQuantum2021}              &  $0.3$ s 			& $0.03f_\text{dec}$ s            & $0.3$ s             \\
\cline{2-5}
$T1_\text{idle}^\text{e}$ & \multicolumn{4}{c|}{$300$ s}             \\
\cline{2-5}
$T2_\text{idle}^\text{n}$ & \multicolumn{4}{c|}{$10$ s}             \\
\cline{2-5}
\vspace{-1.8pt}$T2_\text{link}^\text{n}$ & \vspace{-1.8pt}$0.0075$ s\;\cite{pompiliRealizationMultinodeQuantum2021}    & \vspace{-1.8pt}$0.075$ s 				& $0.0075f_\text{dec}$ s              & \vspace{-1.8pt}$0.075$ s             \\
\cline{2-5}
$T2_\text{idle}^\text{e}$ & \multicolumn{4}{c|}{$1.0$ s}             \\
\cline{2-5}
$t_\text{pulse}$ & \multicolumn{4}{c|}{$1.0\cdot10^{-3}$ s}             \\
\cline{2-5}
$n_\text{DD}$ & 500              & \multicolumn{2}{c|}{18}              & Eq.~\eqref{eq:find_n_DD}            \\
\cline{2-5}
\hline
\multicolumn{5}{c}{} \\ [-5pt]
\multicolumn{5}{c}{\textit{Link efficiency}---see Eq.~\eqref{eq:def_link_efficiency}}                                     \\ [0.5ex]
\hline
$\eta_\mathrm{link}^*$ & $2\cdot10^{-1}$ & $2\cdot10^3$ & $200f_\mathrm{dec}$ & $\eta_\mathrm{link}^*(f_\eta)$            \\
\cline{2-5}
\hline
\multicolumn{5}{c}{} \\ [-5pt]
\multicolumn{5}{c}{\textit{Operation noise}}                                     \\ [0.5ex]
\hline
$p_\text{g}$ & $0.01$            & \multicolumn{3}{c|}{\multirow{2}{*}{Threshold}}             \\
\cline{2-2}
$p_\text{m}$ & $0.01$             & \multicolumn{3}{c|}{}             \\
\hline
\end{tabularx}
\caption{Simulation parameters used. The parameters are introduced in Sec.~\ref{sec:diamond_defect_center_model} and App.~\ref{sec:noise_models}. More details on the values used for scaling parameters $f_\text{dec}$, $f_\phi$ and $f_\eta$, as well as the relations used for $p_\text{EE}(f_\phi)$, $F_\text{link}(f_\phi)$, $\eta_\mathrm{ph}(f_\eta)$, $p_\text{link}(f_\eta)$ and $\eta_\mathrm{link}^*(f_\eta)$, can be found in the captions of the respective figures.}
\label{table:simulation_parameters}
\end{table}

\subsection{Current state-of-the-art parameter set}\label{subsec:state_of_the_art}
Let us first consider a parameter set inspired by state-of-the-art NV center hardware (see the first column of Table~\ref{table:simulation_parameters}). The operation times in this set are based on typical time scales in nitrogen-vacancy centers with a natural $^{13}$C concentration~\cite{Bradley2019, pompiliRealizationMultinodeQuantum2021, bradleyRobustQuantumnetworkMemory2022}---see App.~\ref{sec:experimental_characterization} for more details. The Bell pair parameter values are a collection of the best parameters in current NV center literature---see the first column of Table~\ref{table:simulation_parameters} for relevant citations. In the following, we explicitly refer to this parameter set as the ``state-of-the-art'' parameter set. As discussed in more detail in App.~\ref{sec:experimental_characterization}, for this set, the single-click entanglement protocol outperforms the double-click protocol.

We did not find a noise threshold for the state-of-the-art parameter set---neither with existing GHZ protocols (see Sec.~\ref{subsec:GHZ_protocols_in_literature}) nor when optimization over GHZ protocols (see Sec.~\ref{subsec:dynamic_program_to_generate_GHZ_protocols}). This finding is consistent with earlier investigations with a simplified NV center model.~\cite{bradleyRobustQuantumnetworkMemory2022} We identify two main limitations. The first one is the link efficiency: in this regime, the average entanglement generation times are longer than coherence times during entanglement generation---\textit{i.e.}, $\eta_\textrm{link}^*<1$. 
On top of that, the Bell pair fidelity is relatively low. A low Bell pair fidelity requires complex distillation protocols to achieve high-quality GHZ states. This, in turn, magnifies the impact of decoherence.

\subsection{Near-term parameter sets}\label{subsec:results_near_term}
As expected, further experimental progress and improved fidelities are required for fault-tolerant quantum computation. In the remainder of this section, we characterize two key parameters that drive the code performance in this regime. These findings can be used to guide future hardware development. Specifically, we investigate the effect of improving the Bell pair fidelity and the link efficiency.

\subsubsection{Sensitivity to Bell pair fidelity}\label{subsec:varying_bell_state_fidelity}
Firstly, we investigate the influence of the Bell pair fidelity by using a near-future setting parameter set---see the second column in Table~\ref{table:simulation_parameters}. Compared to the state-of-the-art parameter set of Sec.~\ref{subsec:state_of_the_art}, in this set coherence times during entanglement creation and the photon detection probability are one and two orders of magnitude higher, respectively. The double-click entanglement protocol now gives rise to the best combination of entanglement success probability and Bell pair fidelity, as explained in more detail in App.~\ref{sec:experimental_characterization}. This means that these near-future parameters allow for an increase in the link efficiency by four orders of magnitude compared to the state-of-the-art parameter set of Sec.~\ref{subsec:state_of_the_art}---see Eqs.~\eqref{eq:def_link_efficiency},~\eqref{eq:single_click_model}, and~\eqref{eq:double_click_model}.

\begin{figure}
\centering
\includegraphics[width=\linewidth]{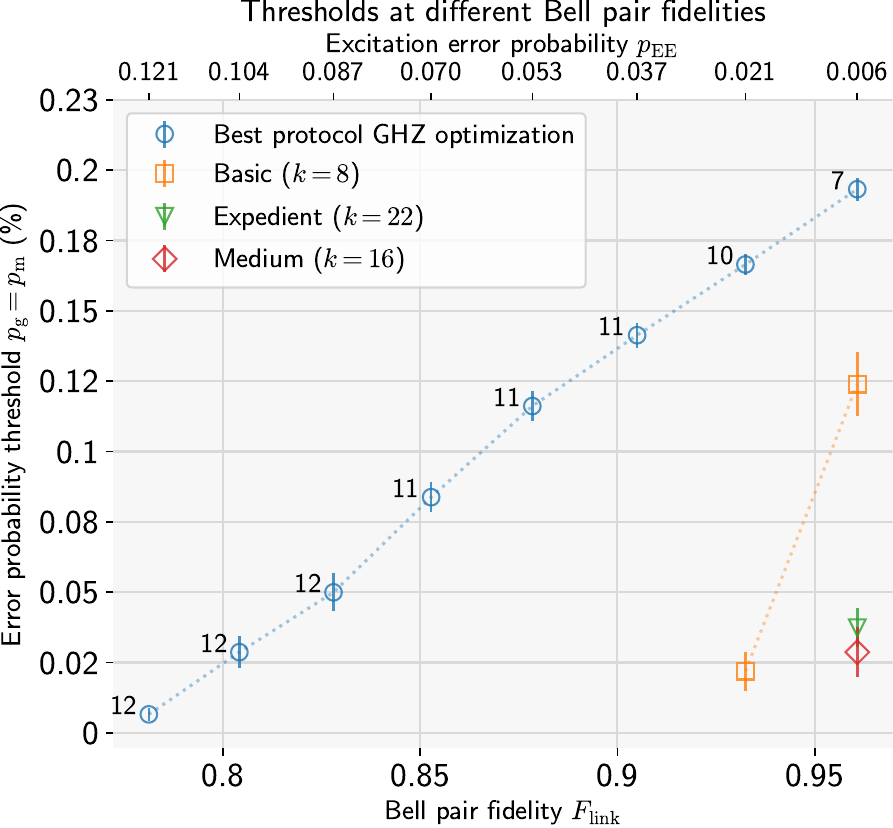}
\caption{Toric surface code error probability thresholds found for $p_\text{g}=p_\text{m}$, at various Bell pair fidelities $F_\text{link}$ (see Table~\ref{table:simulation_parameters} for the parameter values). For all points on the horizontal axis of this plot, we have set $\mu=0.95$ and $F_\text{prep}=0.999$, and varied the excitation error probability $p_\text{EE}$. This leads to different values for the parameter $\phi$ that describes the fidelity of the Bell pairs---see App.~\ref{subsec:bell_state} for more details. For $f_\phi\in\{0, 1, 2, 3, 4, 5, 6, 7, 8\}$, we use $p_\text{EE}(f_\phi)=1-(\phi(f_\phi)/(\sqrt{\mu}(2F_\text{prep}-1)^2))^{1/2}$, with $\phi(f_\phi)=0.72+0.03f_\phi$. In case of similar performance for the best protocols found in the GHZ optimization, we show the protocol with the lowest $k$ value. This value is printed above the blue markers.}
\label{fig:bell_state_quality_plot}
\end{figure}

In our Bell pair model, several parameters contribute to the infidelity of the Bell pair states similarly---\textit{i.e.}, through the parameter $\phi$ of Eq.~\eqref{eq:phi_parameter} that captures all dephasing noise of the model. To investigate the sensitivity of the performance with respect to the Bell pair fidelity, we vary the influence of dephasing by scaling the probability of double excitation probability and off-resonant excitation errors. These are considered one of the leading error sources in present experiments.~\cite{hermansQubitTeleportationNonneighbouring2022} We show the results in Fig.~\ref{fig:bell_state_quality_plot}. In this figure, the bottom of the horizontal axis indicates the Bell pair fidelity; the top indicates the corresponding excitation error probability. 

We find $p_\text{g}=p_\text{m}$ thresholds between $p_\textrm{th}=0.0066(25)$\% for $F_\text{link}\approx0.78$ and $p_\textrm{th}=0.193(4)\%$ for $F_\text{link}\approx0.96$. Interestingly, the minimum fidelity for which we find a threshold, $F_\text{link}\approx0.78$, is lower than the state-of-the-art Bell pair fidelity demonstrated with both the single-click and double-click protocols.~\cite{hensenLoopholefreeBellInequality2015a, hermansQubitTeleportationNonneighbouring2022} This is possible because the link efficiency allows performing several distillation steps. 

We find different optimal protocols as a function of the Bell pair fidelity. In particular, we find that the optimal protocols require more distillation steps as we reduce the Bell pair fidelity, ranging from $k=12$ for $F_\textrm{link}\approx0.78$ to $k=7$ for $F_\textrm{link}\approx0.96$. We find lower thresholds as we decrease the Bell pair fidelity since the more complex distillation protocols amplify the effect of decoherence and require more gates. Furthermore, since existing GHZ creation protocols either have a small number ($k\le8$) or many ($k\ge16$) distillation steps, we can understand why the new protocols with $k\in\{10,11,12\}$ outperform them in this regime.

\subsubsection{Sensitivity to the link efficiency}\label{subsec:varying_link_efficiency}
Secondly, we investigate the influence of the link efficiency for near-future parameter values. In particular, we make use of a Bell pair fidelity $F_\textrm{link}\approx0.95$---close to the highest value in the previous subsection---and we investigate two options for varying the link efficiency.

\begin{figure}
\centering
\includegraphics[width=\linewidth]{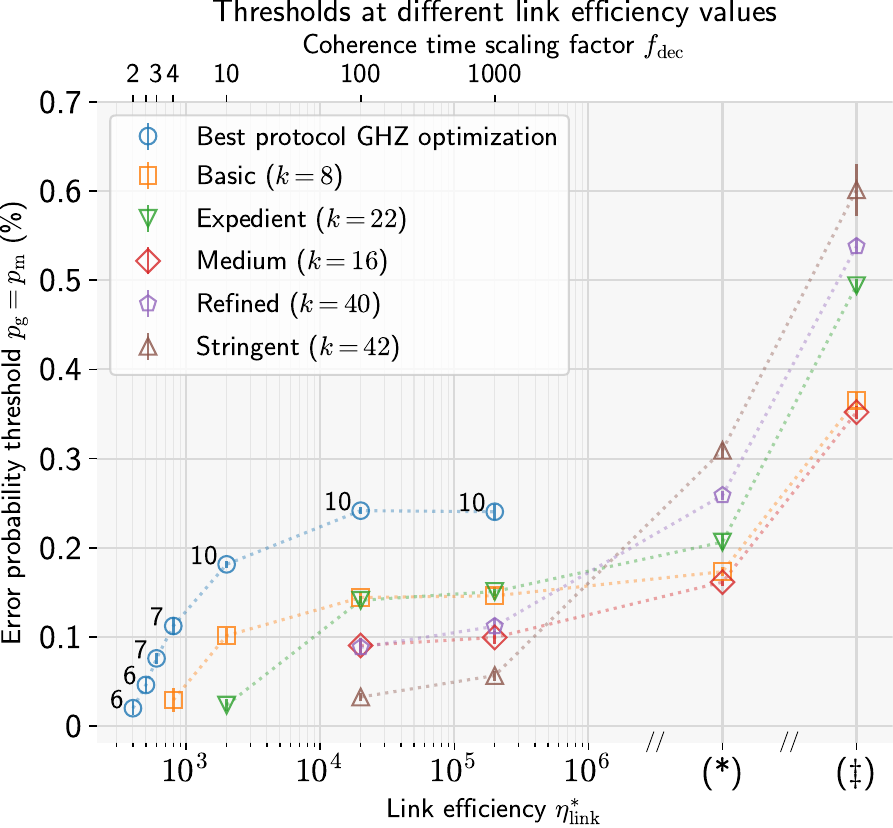}
\caption{Toric surface code error probability thresholds found for $p_\text{g}=p_\text{m}$, at various values of the coherence times during entanglement generation $T1_\text{link}^\text{n}(f_\text{dec})=0.03f_\text{dec}$ seconds and $T2_\text{link}^\text{n}(f_\text{dec})=0.0075f_\text{dec}$ seconds. The $f_\textrm{dec}$ considered are on the top horizontal axis of the plot. The other simulation parameters are in the third column of Table~\ref{table:simulation_parameters}. The corresponding link efficiency is $\eta_\text{link}^*(f_\text{dec})=2f_\text{dec}\cdot10^{2}$. In case of similar performance for the best protocols found in the GHZ optimization, we show the protocol with the lowest $k$ value. This value is printed above the blue markers. Point $(*)$ shows calculations for a scenario without decoherence. Point $(\ddagger)$ shows calculations for a scenario without decoherence and with noiseless SWAP gates.}
\label{fig:link_efficiency_plot}
\end{figure}

First, we vary the link efficiency by varying the coherence times during entanglement generation. For this investigation, which we report in Fig. \ref{fig:link_efficiency_plot}, we use the parameter set of the third column of Table \ref{table:simulation_parameters}. In this set, we use a high photon detection probability $\eta_\mathrm{ph}=0.4472$, leading to an optical entanglement success probability of $p_\textrm{link}=0.1$. 
The $p_\text{g}=p_\text{m}$ threshold values vary between $p_\textrm{th}=0.020(8)$\% with coherence times corresponding to $\eta_\text{link}^*=4\cdot10^2$ and $p_\textrm{th}=0.240(9)$\% for coherence times corresponding to $\eta_\text{link}^*=2\cdot10^5$. 
For coherence times corresponding to $\eta_\textrm{link}^*=3\cdot10^2$ and lower, we did not find thresholds. 
At the other end of the spectrum, we evaluate the thresholds in an idealized modular scenario: in particular, in the absence of decoherence and with perfect SWAP gates (the last two points on the horizontal axis of Fig.~\ref{fig:link_efficiency_plot}). The last point corresponds to a scenario similar to the one analyzed in Ref.~\onlinecite{Nickerson2013}. We report a similar threshold value. For the Stringent protocol, the difference of $p_\textrm{th}=0.775$\% reported in Ref.~\onlinecite{Nickerson2013} and $p_\textrm{th}=0.601(29)$\% found here can be attributed to the choice of the error-syndrome decoder and a reduced number of syndrome measurement cycles.

\begin{figure}
\centering
\includegraphics[width=\linewidth]{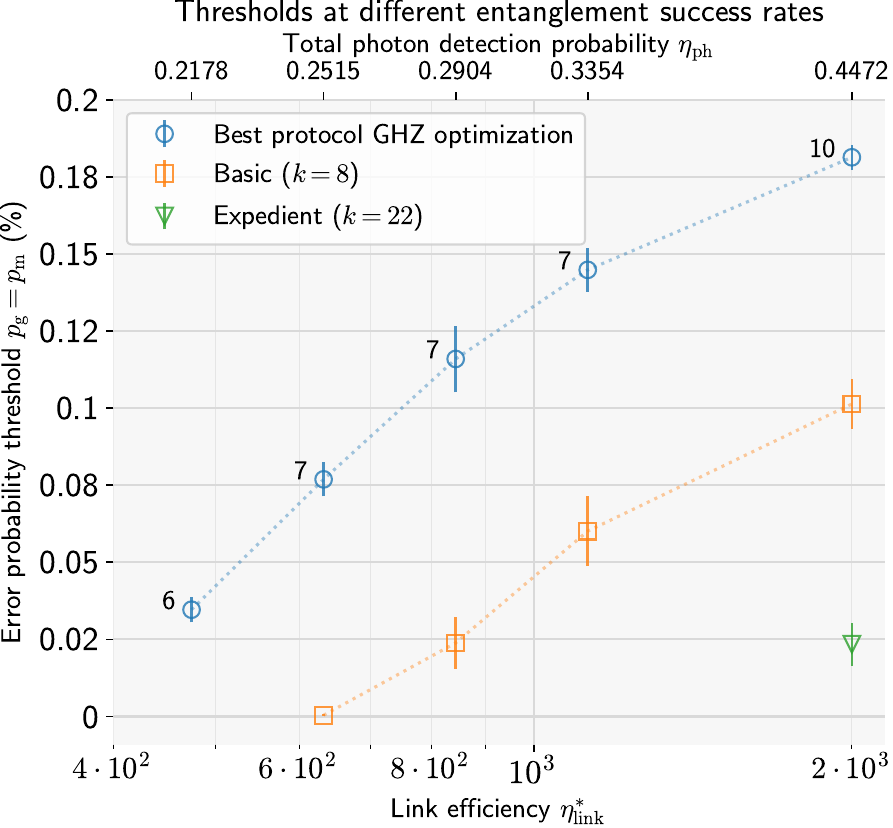}
\caption{Toric surface code error probability thresholds found for $p_\text{g}=p_\text{m}$, at various values of the total photon detection probability $\eta_\mathrm{ph}$. This parameter takes on values $\eta_\mathrm{ph}(f_\eta)=\sqrt{2}\cdot10^{0.0625f_\eta-0.8125}$, for $f_\eta\in\{0,1,2,3,5\}$. With the double-click protocol, this gives rise to $p_\text{link}(f_\eta)=10^{0.125f_\eta-1.625}$ and $\eta_\text{link}^*(f_\eta)=0.002\cdot10^{0.125f_\eta+5.375}$. The other simulation parameters are in the fourth column of Table~\ref{table:simulation_parameters}. In case of similar performance for the best protocols found in the GHZ optimization, we show the protocol with the lowest $k$ value. This value is printed above the blue markers.}
\label{fig:bell_succ_prob_plot}
\end{figure}

We now verify that, in this regime, similar thresholds can instead be found by varying the link efficiency via the entanglement generation rate. Specifically, we vary the entanglement success probability by adjusting the total photon detection probability $\eta_\mathrm{ph}$. For this investigation, which we report in Fig.~\ref{fig:bell_succ_prob_plot}, we use the parameter set in the fourth column of Table~\ref{table:simulation_parameters}. This set contains coherence times during entanglement generation that are ten times higher than the state-of-the-art coherence times of Sec.~\ref{subsec:state_of_the_art}.~\cite{pompiliRealizationMultinodeQuantum2021} We find $p_\text{g}=p_\text{m}$ thresholds between $p_\textrm{th}=0.035(4)$\% for a photon detection probability corresponding to $\eta_\textrm{link}^*\approx4.7\cdot10^2$ and $p_\textrm{th}=0.181(4)$\% for a photon detection probability corresponding to $\eta_\textrm{link}^*=2\cdot10^3$. At photon detection probability corresponding to $\eta_\textrm{link}^*\approx3.6\cdot10^2$ and lower, we are not able to find threshold values. 

The second investigation gives rise to a similar required link efficiency ($\eta_\textrm{link}^*\approx4.7\cdot10^2$) as the first investigation ($\eta_\textrm{link}\approx4\cdot10^2$). The small difference can be attributed to the slightly larger influence of the idling coherence time $T2_\textrm{idle}^\textrm{n}$ in a scenario with a smaller entanglement rate. This shows that the link efficiency captures the key trade-off between cycle duration and decoherence rate, even when experimental overhead such as dynamical decoupling is accounted for. 

We find that the parameter set used determines which GHZ protocol works the best. However, for a large range of parameters close to the state-of-the-art set, one protocol with $k=7$ performs the best. We call this protocol Septimum and detail it in Fig.~\ref{fig:protocol}. In particular, this protocol is (one of) the best-performing protocol(s) at $F_\textrm{link}\approx0.96$ in Fig.~\ref{fig:bell_state_quality_plot}, in the range $5\cdot10^2\lessapprox\eta_\textrm{link}^*\lessapprox2\cdot10^3$ in Fig.~\ref{fig:link_efficiency_plot}, and in the range $6.3\cdot10^2\lessapprox\eta_\textrm{link}^*\lessapprox1.1\cdot10^3$ in Fig.~\ref{fig:bell_succ_prob_plot}. We identify four additional well-performing protocols found with our dynamic program in App.~\ref{app:best_performing_protocols}.

\subsection{GHZ cycle time sensitivity}\label{subsec:cut_off_dependence}
\begin{figure}
\centering
\includegraphics[width=\linewidth]{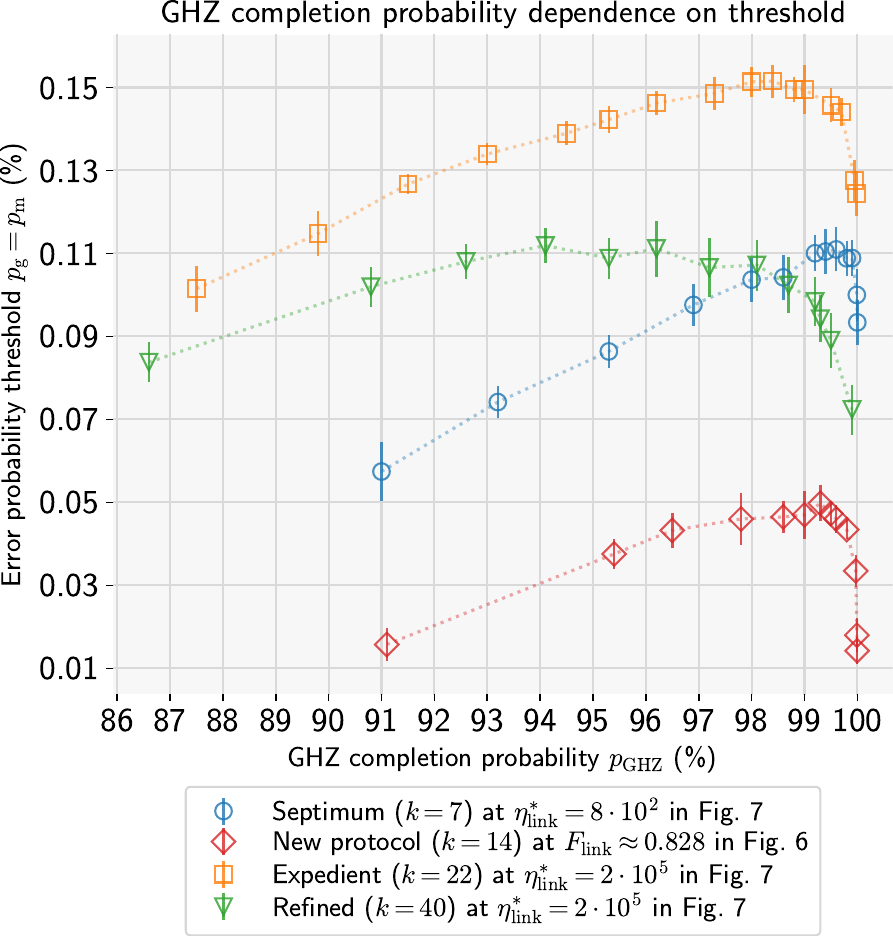}
\caption{Dependence of toric surface code error probability thresholds for $p_\text{g}=p_\text{m}$ on GHZ completion probability $p_\mathrm{GHZ}$. The dependence is plotted for four protocols with a varying number of distillation steps $k$. Each data point is calculated with a different GHZ cycle time $t_\mathrm{GHZ}$. The GHZ completion probability is the probability for a protocol to finish within $t_\mathrm{GHZ}$. In Fig.~\ref{fig:cut_off_dependence_subplots}, we plot the threshold values against the specific $t_\mathrm{GHZ}$ times used to achieve these results.
}
\label{fig:cut_off_time_dependence_plot}
\end{figure}

In the following, we investigate the sensitivity of threshold values to the GHZ cycle time and the associated GHZ completion probability for our diamond defect center hardware model. We present the results in Fig.~\ref{fig:cut_off_time_dependence_plot}.
In this figure, we see a clear dependence of the optimal GHZ completion probability on protocol complexity. In particular, protocols that take longer to finish (\textit{i.e.}, protocols with more distillation steps) peak at lower GHZ completion probabilities than those that finish faster, due to their increased susceptibility to decoherence. We see that for a protocol with relatively small $k$, GHZ cycle times that correspond to GHZ completion probabilities between 99.2 and 99.8\% give rise to the highest threshold values in the parameter regimes considered here, whereas protocols with a large $k$ peak at GHZ completion probabilities between approximately 92.5 and 98.5\%. 

We notice that, for some GHZ protocols, noise thresholds are found at relatively low GHZ completion probabilities of 90\% and lower. This behavior can be directly attributed to the decoder heralding failures in the GHZ generation: as mentioned in Sec.~\ref{subsec:distributed_toric_code_calculations}, we utilize the stabilizer outcome from the previous time layer if a GHZ protocol does not finish within the GHZ cycle time, as opposed to naively performing the stabilizer measurement with the state produced by an unfinished GHZ protocol.

The results in Fig.~\ref{fig:cut_off_time_dependence_plot} show that thresholds can strongly vary on the GHZ cycle time. For computational reasons, except for the results in this subsection, we do not optimize over the GHZ cycle time. Instead, we use a heuristic method to select this time based on the $k$ value of each protocol. We describe this method in App.~\ref{subsec:ghz_cycle_time_settings}.

\section{Discussion: feasibility of parameter sets below threshold}
In the previous section, we observed that the state-of-the-art parameter set is above the threshold. We identified two apparent drivers for this behavior: the Bell pair fidelity and the link efficiency. The sensitivity investigation shows that with a high link efficiency, the requirements on the Bell pair fidelity are modest, while even with a high Bell pair fidelity a high link efficiency is still necessary. 

Let us first discuss the experimental feasibility of the minimum link efficiency $\eta_\textrm{link}^*\gtrapprox4\cdot10^2$  found in Fig.~\ref{fig:bell_succ_prob_plot}. First of all, the link efficiency can be increased by either increasing the coherence times of the data and memory qubits, or by increasing the entanglement success probability---or by a combination of both. In the previous section, we found thresholds with a high success probability ($p_\textrm{link}=0.1$) and a modest increase in the coherence times. However, we also found that with high coherence times during entanglement generation (ten times higher than the state-of-the-art~\cite{pompiliRealizationMultinodeQuantum2021}) and Bell pair fidelities of $F_\mathrm{link}\approx0.95$, the total photon detection probability needs to fulfill $\eta_\mathrm{ph}\gtrapprox0.19$ (Fig.~\ref{fig:bell_succ_prob_plot}). This is a factor fifty above the state-of-the-art parameter value.

The total photon detection probability is the product of multiple probabilities (see App.~\ref{subsec:bell_state}). Present network experiments utilizing NV centers are particularly limited by two of these: the probability of emitting in the zero-phonon-line (ZPL, $\approx3\%$)~\cite{bernienHeraldedEntanglementSolidstate2013} and the total collection efficiency ($\approx10-13\%$).~\cite{hensenLoopholefreeBellInequality2015a,pompiliRealizationMultinodeQuantum2021} Both values are expected to increase by Purcell enhancement of the NV center emission, for example with the use of optical microcavities~\cite{rufResonantExcitationPurcell2021, riedelDeterministicEnhancementCoherent2017} or nanophotonic devices.~\cite{barclayHybridNanocavityResonant2011,liCoherentSpinControl2015} However, even with such devices, the feasibility remains unclear. For microcavities, predicted ZPL emission and collection probabilities are of the order 10 to 46\%\;\cite{rufResonantExcitationPurcell2021, riedelDeterministicEnhancementCoherent2017} and 49\%,~\cite{rozpedekNeartermQuantumrepeaterExperiments2019} respectively. Moreover, the successful integration of Purcell-enhanced and optically coherent NV centers in nanophotonic devices remains an open research challenge due to the detrimental effects of surface charges.~\cite{orphal-kobinOpticallyCoherentNitrogenVacancy2023} 

This realization has led to an increased interest in other color centers in the diamond lattice, as, \textit{e.g.}, SiV and SnV defect centers.~\cite{knautEntanglementNanophotonicQuantum2023} These centers have higher intrinsic emission probabilities into the ZPL---for SnV centers this is reportedly in the area of 60\%,~\cite{gorlitzSpectroscopicInvestigationsNegatively2020} whereas SiV centers approximately emit 70 to 80\% into the ZPL.~\cite{moody2022RoadmapIntegrated2022} Additionally, the inversion symmetry of SnV and SiV centers makes them less susceptible to proximal charges, facilitating integration into nanophotonic devices. Nanophotonic structures offer advantages over microcavities, such as stronger cooperativities enabled by the small mode volumes,~\cite{Sipahigil2016} and reduced sensitivity to vibrations of the cryostat hosting the emitter.~\cite{rufResonantExcitationPurcell2021}
A disadvantage of SnV and SiV centers over NV centers is the fact that they need to be operated at lower temperatures~\cite{sukachevSiliconVacancySpinQubit2017} or under high strain~\cite{sohnControllingCoherenceDiamond2018,stasRobustMultiqubitQuantum2022} to achieve similar coherence times. 

Additionally, these alternative trajectories provide opportunities for ``direct'' GHZ generation schemes, where a GHZ state is created without Bell pair fusion.~\cite{beukersTutorialRemoteEntanglement2023} Contrary to the photon-emission-based Bell pair generation with NV centers, these direct GHZ state generation schemes could be based on the transmission or reflection of photons. Since, for nodes with a single communication qubit, SWAP gates are unavoidable when performing fusion, getting rid of SWAP gates during GHZ state generation could relax the requirements for, \textit{e.g.}, the link efficiency and the photon detection probability.

\section{Conclusion}
In this paper, we investigated the influence of decoherence and other noise sources on a fault-tolerant distributed quantum memory channel with the toric code. 
For this, we developed an open-source package that optimizes GHZ distillation for distributed stabilizer measurements and quantifies the impact of realistic noise sources.~\cite{deboneDataSoftwareUnderlying2024} The GHZ protocols found with this package are compatible with a second open-source package that calculates logical error rates of the (distributed) surface code.~\cite{huOOPSurfaceCode} 

We focused our attention on a specific set of noise models inspired by diamond defect centers. We first observed that state-of-the-art nitrogen-vacancy center hardware does not yet satisfy the thresholds. A parameter-sensitivity analysis shows that the main driver of the performance is the link efficiency, giving a benchmark for future experimental efforts. The photon detection probability of state-of-the-art hardware appears to represent the main challenge for operating the surface code below threshold. Sufficient photon detection probabilities could be achieved with the help of Purcell enhancement of NV center emission, or using other color centers such as silicon-vacancy centers or tin-vacancy centers. The use of other color centers also presents opportunities for schemes that directly generate GHZ states between the communication qubits of more than two nodes---\textit{i.e.}, without fusing Bell pairs.~\cite{beukersTutorialRemoteEntanglement2023} 

With our detailed noise models, we found threshold values up to $0.24$\%. This is three to four times lower than prior thresholds found with less-detailed models. Similarly, the optimal distillation protocols have a small number of distillation steps compared to prior work. For a large parameter regime of parameters, a protocol consuming a minimum of seven Bell pairs was optimal. Its experimental demonstration would be an important step for showing the feasibility of this approach for scalable quantum computation. 

We performed a thorough optimization of GHZ distillation protocols. However, further improvements in other elements of the distributed architecture could partially bridge the gap with the performance of monolithic architectures. For instance, the surface code decoder could model unfinished GHZ distribution rounds as erasure noise. The conversion of binary trees to protocol recipes can be optimized and Bell pair distribution could be scheduled dynamically. On top of that, since our software allows for the implementation of general hardware models, further research could focus on analyzing and understanding a broad range of physical systems in the distributed context. In addition to exploring alternative hardware systems, it would be intriguing to implement a more in-depth model of the system's micro-architecture.

\begin{acknowledgments}
The authors would like to thank Hans Beukers, Johannes Borregaard, Kenneth Goodenough, Fenglei Gu, Ronald Hanson, Sophie Hermans, Stacey Jeffery, Yves van Montfort, Jaco Morits, Siddhant Singh, and Erwin van Zwet for helpful discussions and feedback. We gratefully acknowledge support from the joint research program ``Modular quantum computers'' by Fujitsu Limited and Delft University of Technology, co-funded by the Netherlands Enterprise Agency under project number PPS2007. 
This work was supported by the Netherlands Organization for Scientific Research (NWO/OCW), as part of the Quantum Software Consortium Program under Project 024.003.037/3368. This work was supported by the Netherlands Organisation for Scientific Research (NWO/OCW) through a Vidi grant. This project has received funding from the European Research Council (ERC) under the European Union’s Horizon 2020 research and innovation programme (grant agreement No. 852410). D.E. was partially supported by the JST Moonshot R\&D program under Grant JPMJMS226C. We thank SURF (www.surf.nl) for the support in using the National Supercomputer Snellius.
\end{acknowledgments}

\section*{Author declarations}
\subsection*{Conflict of interest}
The authors have no conflicts to disclose.

\subsection*{Author contributions}
Sébastian de Bone, David Elkouss and Paul Möller designed and conceptualized the study. 
Conor Bradley and Tim Taminiau designed, built, and conducted experiments to characterize color centers. Conor Bradley and Tim Taminiau curated the hardware parameter sets.
Sébastian de Bone built the software to generate and evaluate GHZ protocols. 
Sébastian de Bone and Paul Möller built the software to carry out distributed surface code simulations. 
Sébastian de Bone carried out and analyzed the numerical simulations. 
The manuscript is written by Sébastian de Bone with input and revision from all authors. David Elkouss supervised the project.

\textbf{Sébastian de Bone}: Conceptualization (equal); data curation (lead); formal analysis (lead); methodology (equal); software (equal); visualization (lead); writing - original draft (lead); writing - review and editing (lead). \textbf{Conor Bradley}: Resource (lead); writing - review and editing (supporting). \textbf{David Elkouss}: Conceptualization (equal); formal analysis (supporting); supervision (lead); writing - original draft (supporting); writing - review and editing (supporting). \textbf{Paul Möller}: Conceptualization (equal); methodology (equal); software (equal). \textbf{Tim Taminiau}: Resource (supporting); supervision (supporting); writing - review and editing (supporting). 

\section*{Data availability}
The software and data that supports the findings of this study are openly available in 4TU.ResearchData.~\cite{deboneDataSoftwareUnderlying2024}

\appendix

\section{Experimental characterization}\label{sec:experimental_characterization}
The noise models and parameter values we use in this paper are based on experimental observations by Bradley \textit{et al.}~\cite{Bradley2019, bradleyRobustQuantumnetworkMemory2022} and Pompili \textit{et al.}~\cite{pompiliRealizationMultinodeQuantum2021} Experiments were performed on NV centers at temperatures of 3.7 and 4 K. Time scales for gates are based on microwave and radio-frequency pulses for single-qubit gates on the electron and $^{13}$C qubits, respectively. Time scales for two-qubit gates are based on phase-controlled radiofrequency driving of the carbon spins interleaved with dynamical decoupling sequences on the electron state, following the scheme described by Bradley \textit{et al.}~\cite{Taminiau2014, Bradley2019} 

Characterization of the decoherence model and the associated coherence times was achieved by monitoring the drop in probability of detecting Pauli matrix eigenstates at $t>0$ after preparing the state at $t=0$. For the $T_1$ relaxation time, the $\ket{0}$ and $\ket{1}$ states were used, and the expectation value $\expval{Z}$ of a measurement in the Pauli-$Z$ basis was determined after several delay times---the coherence time then directly follows from the observed exponential drop in the expectation value. With the $\ket{+}$ and $\ket{-}$ states and the expectation $\expval{X}$ for measurement in the Pauli-$X$ basis, and with $\ket{+i}$ and $\ket{-i}$ and $\expval{Y}$, the $T_2$ coherence time of our model could be determined. For these four states, the observed exponential decay $T_\text{dec}$ corresponds to the $T_2$ time of our model via $1/T_2=1/T_1-2/T_\text{dec}$.  

The full model is based on the so-called NV$^-$ state, a spin-1 electron qubit with spin projection quantum number $m_s\in\{-1,0,1\}$. Stochastic ionization can convert the NV$^-$ state to the NV$^0$ state with $m_s\in\{-1/2,1/2\}$. Under the present understanding, this spin state is no longer usable as a qubit due to a fast orbital relaxation process.~\cite{baierOrbitalSpinDynamics2020} Since the electron-spin state is used to control the $^{13}$C spins, ionization accordingly dephases the $^{13}$C states. As such, the coherence times of the NV center memory are currently limited by these ionization effects. Ref.~\onlinecite{bradleyRobustQuantumnetworkMemory2022} proposes a method to mitigate ionization-induced memory dephasing by actively recharging the NV$^{0}$ state. They show ionization and recharging can be performed with minimal loss in fidelity in the state of a $^{13}$C spin in an isotopically purified device. This marks an important avenue for future research. In our model, we do not specifically include ionization, but simply absorb its influence in the $^{13}$C coherence times.

Furthermore, Ref.~\onlinecite{bradleyRobustQuantumnetworkMemory2022} showed that isotopically engineered nitrogen-vacancy devices with a reduced $^{13}$C memory qubit concentration (0.01\%) are able to store a quantum state over 10$^5$ entanglement attempt repetitions. Compared to samples with natural $^{13}$C abundance (1.1\%), the memory qubits have a weaker coupling to their color centers. While this increases coherence times during entanglement generation by several orders of magnitude, it also leads to longer time scales for carrying out gates on the memory qubits and gates between the communication and memory qubits. In natural abundance devices, a quantum state can typically be stored over $10^3$ entanglement attempt repetitions,~\cite{pompiliRealizationMultinodeQuantum2021} while demonstrated single qubit $^{13}$C gates are typically $\approx13$ times faster and two-qubit gates are typically $\approx50$ times faster than the isotopically purified samples.

Nonetheless, in this paper, we assume that the diamond color centers contain a natural abundance of carbon memory qubits (1.1\%). Thus we trade off lower coherence times during entanglement generation for faster gates. This choice was made because it is believed that in future systems entanglement success rates are required to be several orders of magnitude higher than current state-of-the-art. In those regimes, fewer entanglement attempts are required and the influence of decoherence during entanglement generation becomes smaller. It is believed that one then benefits more from having the faster operations that samples with natural concentrations of $^{13}$C atoms offer. 

On top of that, Ref.~\onlinecite{bradleyRobustQuantumnetworkMemory2022} found an idling carbon coherence time of $T2_\text{idle}^\text{n}=10$ seconds, which is too low for running the GHZ creation protocols described in this paper when considering the corresponding gate speeds. This time---comparable to those achieved in natural abundance devices \cite{Bradley2019}---was limited predominantly by other impurities in the diamond, but the expected linear scaling of coherence time with isotopic concentration remains to be demonstrated in future work. In regimes with high entanglement success rates, the time duration of the GHZ creation protocols is almost solely described by the duration of the two-qubit C$Z$, CNOT, C$iY$, and SWAP gates, which are $\approx 50$ slower for the isotopically purified samples. Thus, the operation time of any protocol also takes $\approx 50$ time longer with isotopically purified samples. Equivalent performance for isotopically purified samples would require that $T2_\text{idle}^\text{n}$ also increases by a factor of $50$---\textit{i.e.}, from 10 seconds to 500 seconds. 

Further research into isotopically engineered diamond defect centers is ongoing. This will likely lead to a better understanding of the trade-off between the $^{13}$C concentration and the associated operation times and decoherence rates.

\section{Simulation models and settings}\label{sec:noise_models}
\subsection{Bell pair state}\label{subsec:bell_state}
For generating entanglement, we assume that the diamond color centers make use of either the single-click~\cite{cabrilloCreationEntangledStates1999} or double-click~\cite{barrettEfficientHighfidelityQuantum2005a} Bell pair entanglement protocol. We denote the \emph{bright state} population coefficient for the single-click protocol as $\alpha$. Phase uncertainty originating from a path length difference between the two involved parties is modeled as a dephasing channel on one of the photonic states before the Bell state measurement, with fidelity~\cite{dahlbergLinkLayerProtocol2019, rozpedekNeartermQuantumrepeaterExperiments2019}
\begin{equation}
\lambda=\frac{1}{2}\bigg(1+\frac{I_1(\sigma(\varphi)^{-2})}{I_0(\sigma(\varphi)^{-2})}\bigg).
\end{equation}
Here, $I_0$ and $I_1$ are modified Bessel functions of the zeroth and first order, and $\sigma(\varphi)$ is the standard deviation of the phase instability. We only include phase uncertainty for the single-click protocol, for which a phase instability of 14.3$^\circ$ corresponds to $\lambda=0.984$.~\cite{humphreysDeterministicDeliveryRemote2018} 

The parameter $\eta_\mathrm{ph}$ describes the total photon detection probability per excitation. It can be considered as the product of the total collection efficiency (the transmissivity between defect center and detector multiplied by the detector efficiency) with the probability that a photon is emitted in the detection (time) window and in the zero-phonon line. The photon detection probability only influences the success probability of the entanglement protocol. 

We note that, in our model, we neglect photon detector \emph{dark counts}. This is because we consider regimes in which dark counts are negligible. 

The parameter $p_\text{EE}$ describes the probability of an excitation error during a heralded entanglement generation event. This can occur because an extra photon was emitted during the excitation pulse (a phenomenon known as \emph{double excitation}) or as a result of exciting the dark state. We assume that these excitation errors give rise to a dephasing channel on one of the qubits of the Bell pair. 

For NV centers, the double excitation probability is estimated between 4\% and 7\%.~\cite{humphreysDeterministicDeliveryRemote2018, pompiliRealizationMultinodeQuantum2021} The off-resonant excitations are typically assumed to be negligible. This is because the polarization of the light pulse only leads to a weak driving field on transitions close to the main (bright state) transition, and other transitions are sufficiently far off-resonant. 

For other systems, the situation might be different. Here, one can design the excitation pulse to induce a $\pi$ transition on the main transition and a full $2\pi$ rotation on the closest unwanted transition. Tiurev \textit{et al.} created a model that, based on the energy difference between the main transition and the closest unwanted transition, allows one to estimate $p_\textrm{EE}$---\textit{i.e.}, both the double excitation probability and the probability of exciting this unwanted transition.~\cite{tiurevFidelityTimebinentangledMultiphoton2021, mirambellFidelityCharacterizationSpinphoton2019} Their model shows that, typically, the larger the energy difference is between the two transitions, the smaller $p_\text{EE}$ becomes. 

The single-click Bell pair state $\rho^{(\text{sc})}$ is modeled in the following way:
\begin{equation} \label{eq:single_click_model}
\begin{split}
\rho^{(\text{sc})}&=F_{+}^{(\text{sc})}\ketbra{\Psi^+}{\Psi^+}+F_{-}^{(\text{sc})}\ketbra{\Psi^-}{\Psi^-}\\
&\,\,\,\,\,+\Big(1-F_{+}^{(\text{sc})}-F_{-}^{(\text{sc})}\Big)\ketbra{00}{00},\\
F_{\pm}^{(\text{sc})}&=\frac{1}{p_\text{link}^{(\text{sc})}}(1\pm\phi)\eta_\mathrm{ph}\alpha(1-\alpha),\\
p_\text{link}^{(\text{sc})}&=2\eta_\mathrm{ph}\alpha+\eta_\mathrm{ph}^2\alpha^2\frac{\mu-3}{2}.
\end{split}
\end{equation}
Here, $\ket{\Psi^{\pm}}=(\ket{01}\pm\ket{10})/\sqrt{2}$ are Bell states. $F_\text{link}^{(\text{sc})}\equiv F_{+}^{(\text{sc})}$ denotes the fidelity with respect to the target Bell state $\ket{\Psi^+}$. The parameter $p_\text{link}^{(\text{sc})}$ denotes the success probability of a single attempt with this protocol. Further, $\mu$ is the photon indistinguishability per Bell pair measurement.~\cite{dahlbergLinkLayerProtocol2019} The parameter $\phi$ contains all dephasing contributions of the model:
\begin{equation}\label{eq:phi_parameter}
\phi=\sqrt{\mu}(2F_\text{prep}-1)^2(2\lambda-1)(1-p_\text{EE})^2.
\end{equation}
In deriving these expressions, we have assumed the photon detectors to be non-photon-number resolving.

The dephasing parameter $\phi$ comes back in the expression for the double-click Bell pair state $\rho^{(\text{dc})}$ with success probability $p_\text{link}^{(\text{dc})}$:
\begin{equation} \label{eq:double_click_model}
\begin{split}
\rho^{(\text{dc})}&=F_\text{link}^{(\text{dc})}\ketbra{\Psi^+}{\Psi^+}+(1-F_\text{link}^{(\text{dc})})\ketbra{\Psi^-}{\Psi^-},\\
F_\text{link}^{(\text{dc})}&=\frac{1}{2}(1+\phi^2),\\
p_\text{link}^{(\text{dc})}&=\frac{\eta_\mathrm{ph}^2}{2}.
\end{split}
\end{equation}

Our single-click model combines different elements from NV center single-click models for large-scale networks.~\cite{coopmansNetSquidNETworkSimulator2021, rozpedekNeartermQuantumrepeaterExperiments2019, dahlbergLinkLayerProtocol2019, Kalb2017a, humphreysDeterministicDeliveryRemote2018, pompiliRealizationMultinodeQuantum2021} Our model differs from these models in that we neglect dark counts in the photon detectors. More elaborate versions of the single-click and double-click models can be found in Refs.~\onlinecite{hermansEntanglingRemoteQubits2023, avisRequirementsProcessingnodeQuantum2022a}.

For the parameter value sets used in this paper, as presented in Table~\ref{table:simulation_parameters} of the main text, we use the single-click protocol to give an impression of the optimal Bell pair fidelity and success probability with state-of-the-art NV center parameter values. This is the parameter set identified as $F_\text{prep}=0.99$, $p_\text{EE}=0.04$, $\mu=0.9$, $\lambda=0.984$ and $\eta_\mathrm{ph}=0.0046$. In the simulations performed with near-future parameter values, \textit{i.e.}, the set based on $F_\text{prep}=0.999$, $p_\text{EE}=0.01$, $\mu=0.95$, $\lambda=1$ and $\eta_\mathrm{ph}=0.4472$, we use the double-click protocol model to generate the Bell pair states. 

This is because, for the state-of-the-art parameter set, the single-click protocol is the best option. As can be seen in Eqs.~\eqref{eq:single_click_model} and \eqref{eq:double_click_model}, for a specific set of parameter values, the double-click protocol has a fixed Bell pair fidelity and success probability. For the single-click protocol, on the other hand, the bright space population parameter $\alpha$ allows one to trade in a higher fidelity for a lower success probability, and vice versa. This gives us the option to select $\alpha$ such that the success probability is the same for both protocols. If, for that value of $\alpha$, the state generated by the single-click protocol is better than the state generated with the double-click protocol, one could argue that the single-click protocol is the best choice. 

This is the case for the state-of-the-art parameter set, as we get, with our double-click model, $p_\text{link}^\text{(dc)}\approx1\cdot10^{-5}$ and $F_\text{link}^\text{(dc)}\approx0.852$. With our single-click model, using $\alpha\approx0.00115$, we get a similar success probability, but a state with higher fidelity: $F_\text{link}^\text{(sc)}\approx0.905$. We note that setting $\alpha$ this low is typically not possible in practical situations. However, with the state-of-the-art parameter set, also for higher values of $\alpha$ the single-click model produces better success probabilities and fidelities than the double-click model. In Table~\ref{table:simulation_parameters}, we have used a higher value for $\alpha$, leading to one order of magnitude higher $p_\text{link}^{\text{(sc)}}$, and slightly lower fidelity. 

For the near-future parameters, the double-click model becomes favorable with $p_\mathrm{link}^\mathrm{(dc)}\approx0.1$ and $F_\mathrm{link}^\mathrm{(dc)}\approx0.953$. Setting $\alpha$ to get the same success probability with single-click now only leads to $F_\text{link}^\text{(sc)}\approx0.873$. Technically, for this parameter set, it is possible to reach slightly higher fidelities with single-click than with double-click, but this gives rise to very low success probabilities---\textit{i.e.}, success probabilities unusable for the coherence times considered in this paper.

\subsection{Decoherence}\label{subsec:decoherence}
We describe decoherence noise channels $\mathcal{N}_\text{noise}(\rho)=\sum_{i=1}^{\kappa}K^{(i)}\rho(K^{(i)})^\dagger$ on a general state $\rho$ in terms of their $\kappa$ Kraus operators $\{K^{(i)}\}_{i=1}^\kappa$, with $\sum_{i=1}^{\kappa}K^{(i)}(K^{(i)})^\dagger=\mathbb{I}$.

The generalized amplitude damping channel and dephasing channel used to model NV center decoherence (see Sec.~\ref{sec:diamond_defect_center_model} and Ref.~\onlinecite{Nielsen2000}) make use of the following Kraus operators (with $\gamma_1=1-\exp(-t/T_1)$, where $t$ is the time and $T_1$ is the coherence time):
\begin{equation}
\begin{split}
K_\text{GAD}^{(1)}&=\frac{1}{\sqrt{2}}
\begin{bmatrix}
1 & 0\\
0 & \sqrt{1-\gamma_1}
\end{bmatrix},\\
K_\text{GAD}^{(2)}&=\frac{1}{\sqrt{2}}
\begin{bmatrix}
0 & \sqrt{\gamma_1}\\
0 & 0
\end{bmatrix},\\
K_\text{GAD}^{(3)}&=\frac{1}{\sqrt{2}}
\begin{bmatrix}
\sqrt{1-\gamma_1} & 0\\
0 & 1
\end{bmatrix},\\
K_\text{GAD}^{(4)}&=\frac{1}{\sqrt{2}}
\begin{bmatrix}
0 & 0\\
\sqrt{\gamma_1} & 0
\end{bmatrix}.
\end{split}
\end{equation}

For the phase damping channel, we have the following two Kraus operators (with $\gamma_2=1-\exp(-t/T_2)$, where $T_2$ is the coherence time):
\begin{equation}
K_\text{PD}^{(1)}=
\begin{bmatrix}
1 & 0\\
0 & \sqrt{1-\gamma_2}
\end{bmatrix},\;
K_\text{PD}^{(2)}=
\begin{bmatrix}
0 & 0\\
0 & \sqrt{\gamma_2}
\end{bmatrix}.
\end{equation}

\subsection{Dynamical decoupling sequence length}\label{subsec:dynamical_decoupling_sequence_length}
The coherence times used in our model are based on decoherence in NV center qubits that undergo dynamical decoupling (DD). In Sec.~\ref{sec:diamond_defect_center_model} of the main text, we discuss how DD is implemented in our simulations. 
Each DD sequence has a length of $2n_\text{DD}t_\text{link}+t_\text{pulse}$. The values for $n_\text{DD}$ used in the simulations were obtained by solving the following optimization problem:
\begin{equation}\label{eq:find_n_DD}
n_\text{DD}(p_\text{link})=\min_{n'\in\mathbb{Z}^+}\lim_{\mathcal{A}\rightarrow\infty}\sum_{i=1}^{\mathcal{A}}\sum_{j=1}^{\mathcal{A}}p^{(i)}_\text{link}p^{(j)}_\text{link}t^{(i,j)}_{n'}.
\end{equation}
In Eq.~\eqref{eq:find_n_DD}, we perform the minimization for $n'$ as a member of the positive integers $\mathbb{Z}^+$. The goal is to minimize $n'$ over the average completion time of generating two Bell pairs in parallel, where we also wait for both nodes to finish their DD sequences. Here, $p_\text{link}^{(i)}=p_\text{link}(1-p_\text{link})^{i-1}$ and $p_\text{link}^{(j)}=p_\text{link}(1-p_\text{link})^{j-1}$ denote the probabilities of obtaining entanglement generation success at exactly the $i$th and $j$th attempt. Further, $t_{n'}^{(i,j)}=\lceil \max(i,j)/(2n')\rceil(2n't_\text{link}+t_\text{pulse})$ is the effective time of performing the required entanglement attempts in this specific scenario, where $\lceil \max(i,j)/(2n')\rceil$ describes how many DD sequences are required for these attempts and $2n't_\text{link}+t_\text{pulse}$ describes the time of one DD sequence. To solve Eq.~\eqref{eq:find_n_DD} in a practical setting, it suffices to take a large number for $\mathcal{A}$ instead of letting it go to infinity. Because finding $n_\text{DD}$ in this way only minimizes the waiting and refocusing time during entanglement generation in two nodes, and not the waiting time during the other operations, this process does not lead to the optimal $n_\text{DD}$. We found that it does, however, give rise to values for $n_\text{DD}$ that, typically, produce good results.  

\subsection{Gate and measurement noise}\label{subsec:gate_and_measurement_noise}
For gates, we consider a gate set consisting of the Pauli gates $(X,Y,Z)$, the Hadamard gate, the CNOT (C$X$) gate, the C$Z$ gate, and the C$iY$ gate. These are not the native gates of the NV center, but their true gate set can be compiled into the gate set used here without additional costs in terms of two-qubit gates.~\cite{abobeihFaulttolerantOperationLogical2022} Noise on two-qubit gates are modeled with a depolarizing noise model:
\begin{equation}
\mathcal{N}_\text{g}(\rho)=(1-p_\text{g})\rho+\frac{p_\text{g}}{15}\sum_{(P_i,P_j)}(P_i\otimes P_j)\rho(P_i\otimes P_j)^\dagger,
\end{equation}
where the sum is over $(P_i,P_j)\in\{\mathbb{I},X,Y,Z\}^2\backslash(\mathbb{I},\mathbb{I})$. 

Measurements are restricted to measuring (single-qubit) electron qubits in the Pauli-$Z$ basis. Measuring in the Pauli-$X$ basis is achieved with an additional Hadamard gate. Measurement errors are modeled by a probability $p_\text{m}$ that the measurement projects onto the opposite eigenstate of the measured operator.

\subsection{GHZ cycle time settings}\label{subsec:ghz_cycle_time_settings}
In Sec.~\ref{subsec:cut_off_dependence}, we discuss the influence of the GHZ cycle time $t_\mathrm{GHZ}$ on the surface code threshold value. In this section, we discuss the heuristic method we use for finding a suitable GHZ cycle time for a specific protocol at a specific set of error probabilities. 

\begin{figure*}
    \centering
    \includegraphics[width=\linewidth]{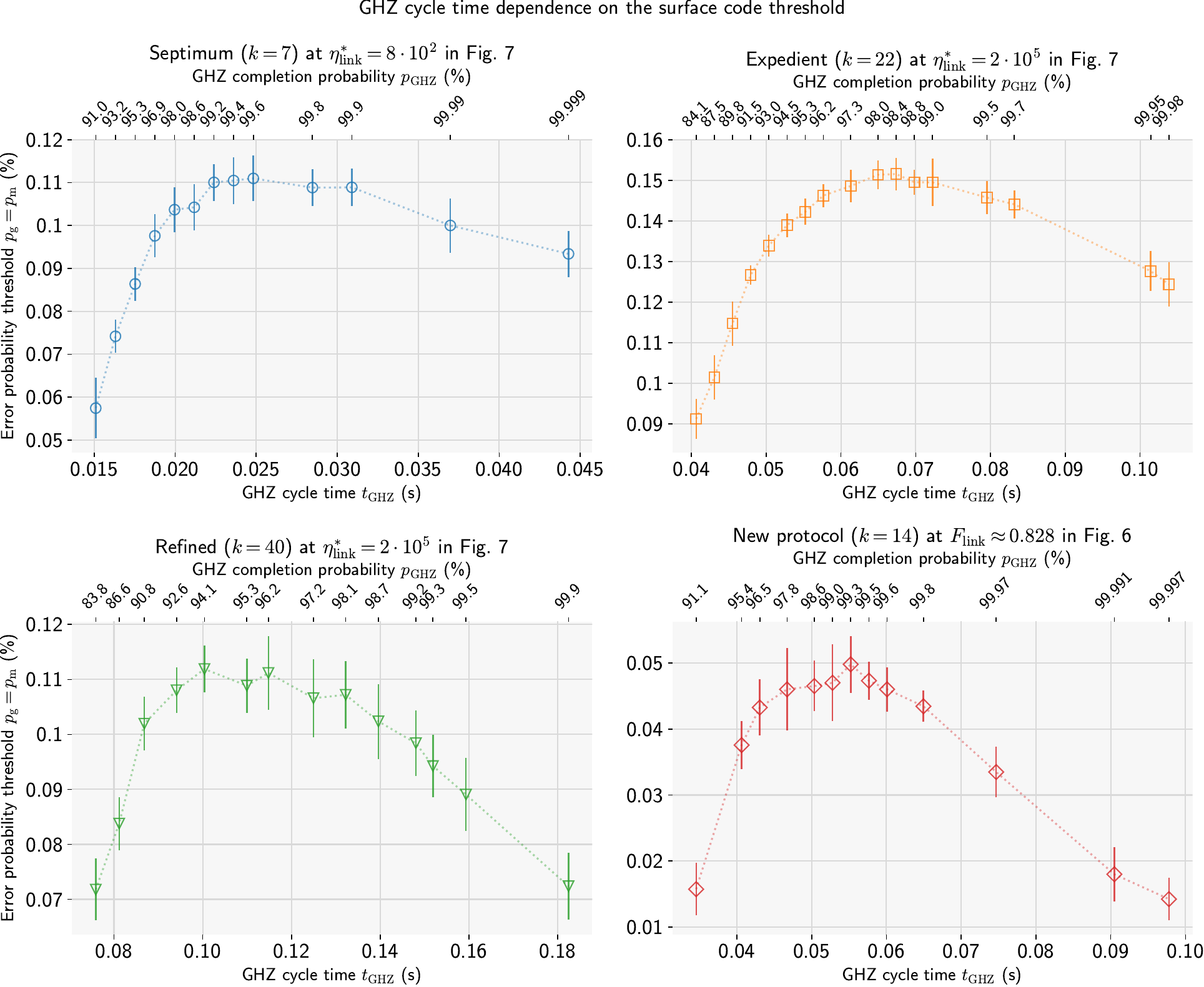}
\caption{Dependence of surface code error probability thresholds for $p_\mathrm{g}=p_\mathrm{m}$ on GHZ cycle time $t_\mathrm{GHZ}$. Each $t_\mathrm{GHZ}$ gives rise to probability $p_\mathrm{GHZ}$ that a protocol has to finish within $t_\mathrm{GHZ}$: for each $t_\mathrm{GHZ}$, the associated probabilities are printed on the top $x$-axis of each plot. In Fig.~\ref{fig:cut_off_time_dependence_plot}, we directly plot the threshold values against $p_\mathrm{GHZ}$ for the same four protocols. 
}
\label{fig:cut_off_dependence_subplots}
\end{figure*}

As discussed earlier, protocols with more distillation steps $k$ take (on average) longer to finish. This means that, compared to a protocol with a smaller $k$, they require a longer $t_\mathrm{GHZ}$ to reach the same GHZ completion probability $p_\mathrm{GHZ}$. However, because decoherence plays a larger role at a higher $t_\mathrm{GHZ}$, we typically see that the threshold values obtained with protocols with higher $k$ peak at a lower GHZ completion probability. This becomes clear in Figs.~\ref{fig:cut_off_time_dependence_plot} and \ref{fig:cut_off_dependence_subplots}, where we plot how the threshold values change when using different GHZ cycle times for four protocols with varying $k$. 

In an attempt to limit calculation times during our GHZ protocol search, we made use of a heuristic-driven method to estimate the optimal $t_\mathrm{GHZ}$ at a certain error probability configuration. Using the protocol-specific parameter $k$, we first identify an adequate GHZ completion probability as $p_\mathrm{GHZ}^\mathrm{(aim)}=(100.2-k/10)\%$. On top of that, we use prior knowledge for a rough estimate $p_\mathrm{th}^\mathrm{(est)}$ of the value of the threshold. We then determine the distribution of the protocol's duration at the error probability $p_\mathrm{th}^\mathrm{(est)}$, by running it without a GHZ cycle time. Finally, using this distribution, we determine $t_\mathrm{GHZ}$ by selecting a time at which \emph{at least} a fraction $p_\mathrm{GHZ}^\mathrm{(aim)}$ of the iterations will finish. 

\section{Details regarding the calculation of the superoperator}\label{app:superoperator_calculations}
\subsection{Superoperator calculation}\label{subsec:superoperator_calculation}
In this section, we describe how we calculate the superoperator that we use in the surface code simulations. Separately calculating a superoperator has the advantage that it breaks up the process of calculating GHZ state creation from the threshold simulations: this drastically decreases the complexity of the full calculation. 

Following earlier work by Nickerson \textit{et al.}, we assume that only Pauli errors occur on the data qubits during the toric code simulations.~\cite{Nickerson2013, nickersonFreelyScalableQuantum2014} This simplifies the simulation, as every stabilizer measurement now deterministically measures either $+1$ or $-1$, and measurement results can be calculated by simply considering commutativity between Pauli errors and the stabilizer operators. In most situations, the stochastic Pauli error model can be considered a good approximation for coherent errors described by continuous rotations.~\cite{greenbaumModelingCoherentErrors2017} On top of that, since the nuclear spin qubits (\textit{i.e.}, the memory qubits) of NV centers have no states to leak to, it is believed that a depolarizing channel (\textit{i.e.}, Pauli noise) is a good approximation for noise on these qubits.

Our characterization of the toric code stabilizer measurements is carried out with density matrix calculations that \emph{do} include more general errors. To align these calculations with the toric code calculations themselves, the stabilizer measurement channel is twirled over the Pauli group.~\cite{durStandardFormsNoisy2005, gellerEfficientErrorModels2013, caiConstructingSmallerPauli2019} 
This makes sure the superoperators describing the channel only contain Pauli errors. Each superoperator is constructed via the channel's Choi state---\textit{i.e.}, by using the GHZ state created by the concerning protocol to non-locally perform the stabilizer measurement on half of the maximally entangled state. 

To explain this process in more detail, we consider the states $\ket{\Psi_\pm}$ that follow from projecting half of the maximally entangled state $\ket{\Psi}$ on the $+P^{\otimes4}$ and $-P^{\otimes4}$ subspaces with projectors $\Pi_\pm=(\mathbb{I}^{\otimes8}\pm P^{\otimes4}\otimes\mathbb{I}^{\otimes4})/2$. Here, $P\in\{X,Z\}$ describes the two types of stabilizer measurements of the toric code, and $\ket{\Psi}$ is the eight-qubit maximally entangled state describing the four data qubits of the code. We also define states $|\Psi_\pm^{(m)}\rangle$ that describe Pauli errors $P_m\in\{\mathbb{I},X,Y,Z\}^{\otimes4}$ occuring on the first half of $\ket{\Psi_\pm}$ after the projection with $\Pi_\pm$. We define:
\begin{equation}
\begin{split}
&\ket{\Psi}=\frac{1}{\sqrt{2^4}}\sum_{j=0}^{2^4-1}\ket{j}\otimes\ket{j},\\
&\ket{\Psi_\pm}=\frac{\mathbb{I}^{\otimes8}\pm P^{\otimes4}\otimes\mathbb{I}^{\otimes4}}{\sqrt{2}}\ket{\Psi},\\
&|\Psi_\pm^{(m)}\rangle=(P_m\otimes\mathbb{I}^{\otimes4})\ket{\Psi_\pm}.
\end{split}
\end{equation}
Later in the analysis, we only consider the subset of Pauli operators $P_m$ that lead to orthogonal states $|\Psi_\pm^{(m)}\rangle$, \textit{i.e.}, we only use $P_m$ that make sure we have 
\begin{equation}
\braket{\Psi_{s}^{(m)}}{\Psi_{s'}^{(n)}}=\delta_{mn}\delta_{ss'},
\end{equation}
with $(s,s')\in\{+,-\}^2$. We call this subset $\mathcal{E}\subseteq\{\mathbb{I},X,Y,Z\}^{\otimes4}$.

We define two versions of the full noisy stabilizer measurement channel: $\mathcal{N}_+$, which projects with $\Pi_+$, and $\mathcal{N}_-$, which projects with $\Pi_-$:
\begin{equation}
\mathcal{N}_s(\rho)=\sum_i K^{(i)}_s\Pi_s\rho \Pi_s(K^{(i)}_s)^\dagger.
\end{equation}
Here, $s\in\{+,-\}$. The Kraus operators $K^{(i)}_s$ describe the noise on the data qubits. Each of them can be decomposed into Pauli matrices:
\begin{equation}
K^{(i)}_s=\sum_{P_q\in\{\mathbb{I},X,Y,Z\}^{\otimes4}}\xi_{s,q}^{(i)}P_q.
\end{equation}
Using this decomposition, the channel's Choi state can be expressed in the following way:
\begin{equation}
\begin{split}
\rho_\text{Choi}&=\sum_s(\mathcal{N}_s\otimes\mathbb{I}^{\otimes4})(\ketbra{\Psi}{\Psi})\\
&=\sum_s\sum_i\sum_{P_q,P_{q'}}\xi_{s,q}^{(i)}\big(\xi_{s,q'}^{(i)}\big)^*\ketbra{\Psi_s^{(q)}}{\Psi_s^{(q')}}.
\end{split}
\end{equation}
We now focus on $\rho_\text{Choi}$ as the post-measurement state for stabilizer measurement outcome $+1$. The influence of noise can cause measurement errors, meaning $\rho_\text{Choi}$ can contain terms projected with $\Pi_-$. One can extract coefficients $p_s^{(m)}$ from $\rho_\text{Choi}$ by constructing the states $|\Psi_s^{(m)}\rangle$ from Pauli operators $P_m\in\mathcal{E}$ via:
\begin{equation}\label{eq:superoperator_elements}
p_s^{(m)}=\expval{\rho_\text{Choi}}{\Psi_s^{(m)}}=\sum_i \abs{\xi_{s,m}^{(i)}}^2.
\end{equation}
These are the coefficients of the Pauli operators that act as the stabilizer measurement channel's Kraus operators after the channel is twirled over the Pauli group. 

We see that this procedure gives us the probabilities required to construct the superoperator of the channel. If $\rho_\text{Choi}$ is constructed by preparing the post-measurement state according to a $+1$ measurement outcome, the coefficients $p_+^{(m)}$ give rise to the Pauli errors $P_m\in\mathcal{E}$ without a measurement error on the stabilizer measurement, whereas the coefficients $p_-^{(m)}$ describe Pauli errors $P_m$ accompanied with a measurement error. If $\rho_\text{Choi}$ is constructed with a $-1$ measurement outcome, the role of $p_+^{(m)}$ and $p_-^{(m)}$ is inverted, but the parameter values themselves are the same.

The stabilizer fidelity is defined as the coefficient $p_s^{(m)}$ corresponding to $P_m=\mathbb{I}^{\otimes4}\otimes\mathbb{I}^{\otimes4}$ (\textit{i.e.}, no errors on the data qubits) and no stabilizer measurement error. In our search for well-performing GHZ creation protocols, a good reason for comparing two protocols by using the stabilizer fidelity over, \emph{e.g.}, the GHZ state fidelity is the fact that the surface code data qubits undergo more decoherence for protocols that take longer to finish. This aspect of the optimization problem is not taken into account if we just use the GHZ fidelity to compare the protocols.

\subsection{Convergence of the average superoperator}
\begin{figure}
    \centering
    \includegraphics[width=\linewidth]{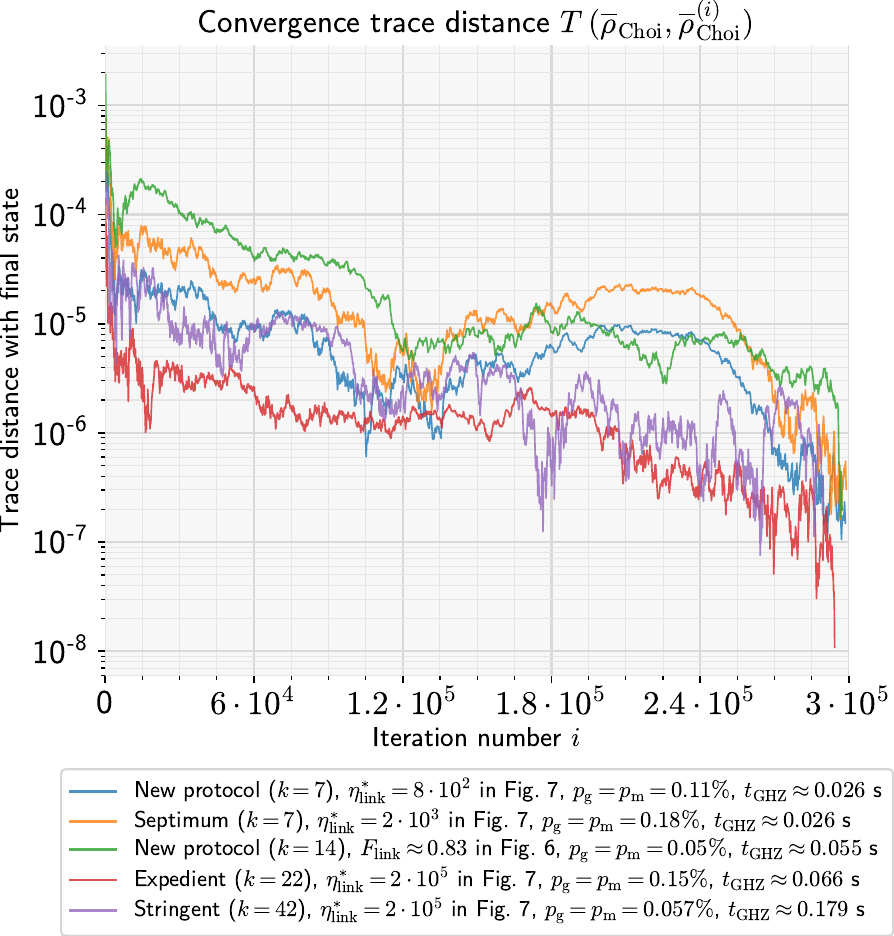}
\caption{Convergence of the trace distance between $\overline{\rho}_\mathrm{Choi}$ (the average Choi state after $3\cdot10^5$ Monte Carlo iterations) and $\overline{\rho}_\mathrm{Choi}^{(i)}$ (the average Choi state after $i$ iterations). We track changes in the trace distances by varying $i$ on the $x$-axis of the plot. For each data point, we add on the order of 100 new iterations to $\overline{\rho}_\mathrm{Choi}^{(i)}$. In the plot, we include five GHZ protocols with a varying number of distillation steps $k$. We only include Choi states based on $X^{\otimes4}$ stabilizer measurements and exclude iterations that did not finish within the GHZ cycle time $t_\mathrm{GHZ}$. 
}
\label{fig:trace_distance_convergence}
\end{figure}

The construction of a superoperator that we use in the surface code simulator requires averaging over a large number of Monte Carlo simulations. In this section, we investigate the convergence of the average superoperator over an increasing number of Monte Carlo samples. In Fig.~\ref{fig:trace_distance_convergence}, we calculate the average Choi state $\overline{\rho}_\mathrm{Choi}$ over $3\cdot10^5$ Monte Carlo iterations and calculate the \emph{trace distance} of this state with the average Choi state $\overline{\rho}_\mathrm{Choi}^{(i)}$ after a smaller number of $i$ iterations. As explained in more detail below, this figure suggests that, after $10^5$ Monte Carlo samples, errors in the average superoperator elements are on the order of $10^{-4}$.

For $s\in\{+,-\}$ and $P_m\in\mathcal{E}$, the superoperator used in threshold simulations is calculated as the average $\overline{S}=\{\overline{p}_s^{(m)}\}_{s,m}$ of the individual superoperators $S=\{p_s^{(m)}\}_{s,m}$ calculated with the method of App.~\ref{subsec:superoperator_calculation}. Alternatively, this average superoperator can be constructed by calculating the average Choi state $\overline{\rho}_\mathrm{Choi}$, as defined in App.~\ref{subsec:superoperator_calculation}, and using Eq.~\eqref{eq:superoperator_elements} to calculate $\{\overline{p}_s^{(m)}\}_{s,m}$ from this average Choi state. 

The trace distance between two density matrices $\rho$ and $\rho'$ is defined as
\begin{equation}
    T(\rho,\rho')=\frac{1}{2}\Tr\sqrt{(\rho-\rho')^\dagger(\rho-\rho')}.
\end{equation}
For an operator $\mathcal{P}$ with eigenvalues $0\leq\xi_j\leq1$, we can show that the following holds:~\cite{Nielsen2000}
\begin{equation}
    \abs{\Tr(\mathcal{P}(\rho-\rho'))}
    \leq T(\rho,\rho').
\end{equation}
This means that, for $\overline{\rho}_\mathrm{Choi}^{(i)}$ after a certain iteration $i$, $T(\overline{\rho}_\mathrm{Choi},\overline{\rho}_\mathrm{Choi}^{(i)})$ provides an upper bound on the difference between the average superoperator $\overline{S}^{(i)}$ after iteration $i$ and the average superoperator $\overline{S}$ after the full number of iterations. This is because, for the difference in the elements of $\overline{S}$ and $\overline{S}^{(i)}$, the following holds:
\begin{equation}
\begin{split}
\Delta \overline{p}_s^{(m)}&=\abs{\expval{\overline{\rho}_\text{Choi}}{\Psi_s^{(m)}}-\expval{\overline{\rho}_\text{Choi}^{(i)}}{\Psi_s^{(m)}}}\\
&=\abs{\Tr\left(\ketbra{\Psi_s^{(m)}}{\Psi_s^{(m)}}\left(\overline{\rho}_\mathrm{Choi}-\overline{\rho}_\mathrm{Choi}^{(i)}\right)\right)}\\
&\leq T\left(\overline{\rho}_\mathrm{Choi},\overline{\rho}_\mathrm{Choi}^{(i)}\right).
\end{split}
\end{equation}
Calculating $T(\overline{\rho}_\mathrm{Choi},\overline{\rho}_\mathrm{Choi}^{(i)})$, therefore, gives us information about the convergence of the average superoperator elements after $i$ iterations. 

In our simulations, as shown schematically in Fig.~\ref{fig:calculation_process}, the superoperator $\{\overline{p}_s^{(m)}\}_{s,m}$ is subsequently used in a second level of Monte Carlo simulations that emulates the operation of the surface code with this specific superoperator. In App.~\ref{app:fitting_procedure_threshold_plots}, we discuss how, for a specific $\{\overline{p}_s^{(m)}\}_{s,m}$, the statistical uncertainty in the Monte Carlo simulations of the surface code leads to uncertainty in the calculated threshold value.

\section{Fitting procedure for determining thresholds}\label{app:fitting_procedure_threshold_plots}
In this appendix, we describe the fitting procedure we use to calculate toric surface code thresholds as well as the associated uncertainties in the threshold values. To perform the fitting procedure described below, we make use of the \code{optimize.curve\textunderscore fit} function of the SciPy package for Python.~\cite{virtanenSciPyFundamentalAlgorithms}

\subsection{Regression model}
Calculating threshold values involves varying one or more of the error probabilities.
For each combination of error probabilities $p$, we calculate an average superoperator using the methods described in App.~\ref{app:superoperator_calculations}. At a specific $p$, we then use Monte Carlo simulations to calculate the logical success rate $r$ of the toric surface code for multiple lattice sizes $L$. 

We denote the observed logical success rate of a certain input combination $(p_i,L_i)$ by $r_i$. We make use of $n_\text{C}$ to describe the total number of input combinations: $\{(p_i,L_i)\}_{i=1}^{n_\text{C}}$. For a single $(p_i,L_i)$ combination, the logical success rate is defined as the number of error-correction iterations $M_i$ that do not induce a logical error divided by the full number of error-correction iterations $N_i$. In the context of this paper, $N_i$ can be considered as the number of Monte Carlo iterations used for surface code calculations for a certain $(p_i,L_i)$ and the exact hardware configuration used. We assume that the uncertainty in the observed logical success rates is described by the binomial distribution. This means that the standard deviation can be estimated via
\begin{equation}
\sigma_i=\sqrt{\frac{r_i(1-r_i)}{N_i}}, \;\;\text{where}\; r_i=\frac{M_i}{N_i}.
\end{equation}

Following Wang~\textit{et al.}, we fit the logical success rates with the following model:~\cite{wangConfinementHiggsTransitionDisordered2003}
\begin{equation}\label{eq:fitting_model_equation}
\hat{r}=\hat{a}+\hat{b}(p-\hat{p}_\text{th})L^{1/\hat{\kappa}}+\hat{c}(p-\hat{p}_\text{th})^2L^{2/\hat{\kappa}}+\hat{d}L^{-1/\hat{\zeta}}.
\end{equation}
Using this model, we find \emph{estimates} $\{\hat{r_i}\}_i$ of the logical success rates for all input combinations $\{(p_i,L_i)\}_i$. For a certain $(p_i,L_i)$, the \emph{residual} $\hat{\epsilon}_i$ is defined as the difference between the observed logical success rate and the estimated value: $\hat{\epsilon}_i=r_i-\hat{r}_i$. Values for the seven fitting parameters $\hat{a}$, $\hat{b}$, $\hat{c}$, $\hat{d}$, $\hat{p}_\text{th}$, $\hat{\kappa}$ and $\hat{\zeta}$ are found by identifying their (local) minimum with respect to the sum $Q$ of the ``weighted'' squared residuals. This sum is defined in the following way:
\begin{equation}\label{eq:q_parameter}
Q=\sum_{i=1}^{n_\text{C}}\bigg(\frac{\hat{\epsilon}_i}{\sigma_i}\bigg)^2=\sum_{i=1}^{n_\text{C}}\bigg(\frac{r_i-\hat{r}_i}{\sigma_i}\bigg)^2.
\end{equation}
We see that this approach makes sure that residuals of data points that are determined with high uncertainty (\textit{i.e.}, with a high standard deviation $\sigma_i$) are given less priority in the least-squares fit than data points with low uncertainty. 

\subsection{Weighted least-squares fitting procedure}
To understand how the confidence intervals in the values of the fitting parameters are determined, we delve a bit deeper into how one could determine fitting parameters for a non-linear regression like Eq.~\eqref{eq:fitting_model_equation}. In line with convention, we denote our input configuration as $X_i=(p_i,L_i)$, and we write $\hat{r}_i$ as $\hat{r}_i=f(\hat{\beta},X_i)$. Here, the function $f$ is the function of Eq.~\eqref{eq:fitting_model_equation} and $\hat{\beta}=(\hat{a},\hat{b},\hat{c},\hat{d},\hat{p}_\text{th},\hat{\kappa},\hat{\zeta})$ describes the converged values for the fitting parameters after the optimization. We define $n_\mathrm{P}=7$ as the number of fitting parameters of the model. Furthermore, we use the notation $\beta^{(t)}$ to indicate the fitting parameter values at a certain step $t$ during the optimization. We can now write $r_i=f(\beta^{(t)},X_i)+\epsilon_i^{(t)}$ for each $t$. Here, the residuals also contain the superscript $(t)$ to denote that they depend on the exact values of $\beta^{(t)}$. 

Finding the least-squares fit is now achieved with the \emph{Gauss-Newton algorithm}.~\cite{ruckstuhlIntroductionNonlinearRegression2010, constalesChapterExperimentalData2017} This method is a variant of Newton's method for finding the minimum of a non-linear function. We start with a guess $\beta^{(1)}$ for the fitting parameter values. These values are then iteratively updated by using the fact that we want to minimize the parameter $Q$ of Eq.~\eqref{eq:q_parameter} until they converge. To go from a certain $\beta^{(t)}$ to a new improved version $\beta^{(t+1)}$, one makes use of $r_i=f(\beta^{(t+1)},X_i)+\epsilon_i^{(t+1)}$ to construct a new estimation in terms of the old fitting parameter values $\beta^{(t)}$. More explicitly, $f(\beta^{(t+1)},X_i)$ is Taylor expanded around $\beta^{(t)}$, and the second and higher order terms are neglected. This allows one to write $r_i=f(\beta^{(t+1)},X_i)+\epsilon_i^{(t+1)}$ as a matrix equation---\textit{i.e.}, with each of the $n_\mathrm{C}$ input and output combinations of $X_i$ and $r_i$ in a separate row of this equation. To this end, we put the fitting parameter values at step $t$ of the optimization process into a $n_\text{P}\times1$ column vector
\begin{equation}
    \boldsymbol{\beta}^{(t)}\equiv[\begin{matrix}a^{(t)}&b^{(t)}&c^{(t)}&d^{(t)}&p_\text{th}^{(t)}&\kappa^{(t)}&\zeta^{(t)}\end{matrix}]^T.
\end{equation}
We do the same for the values of $\{X_i\}_i$ and $\{\epsilon_i^{(t)}\}_i$, and write them as $n_\text{C}\times1$ column vectors $\boldsymbol{X}$ and $\boldsymbol{\epsilon}^{(t)}$, respectively. Finally, we define $\boldsymbol{\Delta\beta}^{(t+1)}$ as $\boldsymbol{\Delta\beta}^{(t+1)}\equiv\boldsymbol{\beta}^{(t+1)}-\boldsymbol{\beta}^{(t)}$. The full (Taylor expanded) version of $r_i=f(\beta^{(t+1)},X_i)+\epsilon_i^{(t+1)}$ can now be rewritten as~\cite{ruckstuhlIntroductionNonlinearRegression2010, constalesChapterExperimentalData2017}
\begin{equation}\label{eq:matrix_equation_residuals}
    \boldsymbol{\epsilon}^{(t+1)}=\boldsymbol{\epsilon}^{(t)}-\boldsymbol{J}^{(t)}\boldsymbol{\Delta\beta}^{(t+1)}.
\end{equation}
Here, $\boldsymbol{J}^{(t)}$ is an $n_\text{C}\times n_\text{P}$ matrix that contains the derivatives of $f$ with respect to the fitting parameters, evaluated at $\beta^{(t)}$ and the inputs $\{X_i\}_i$. This matrix is also known as the \emph{Jacobian} matrix.

\begin{figure}
    \centering
    \includegraphics[width=\linewidth]{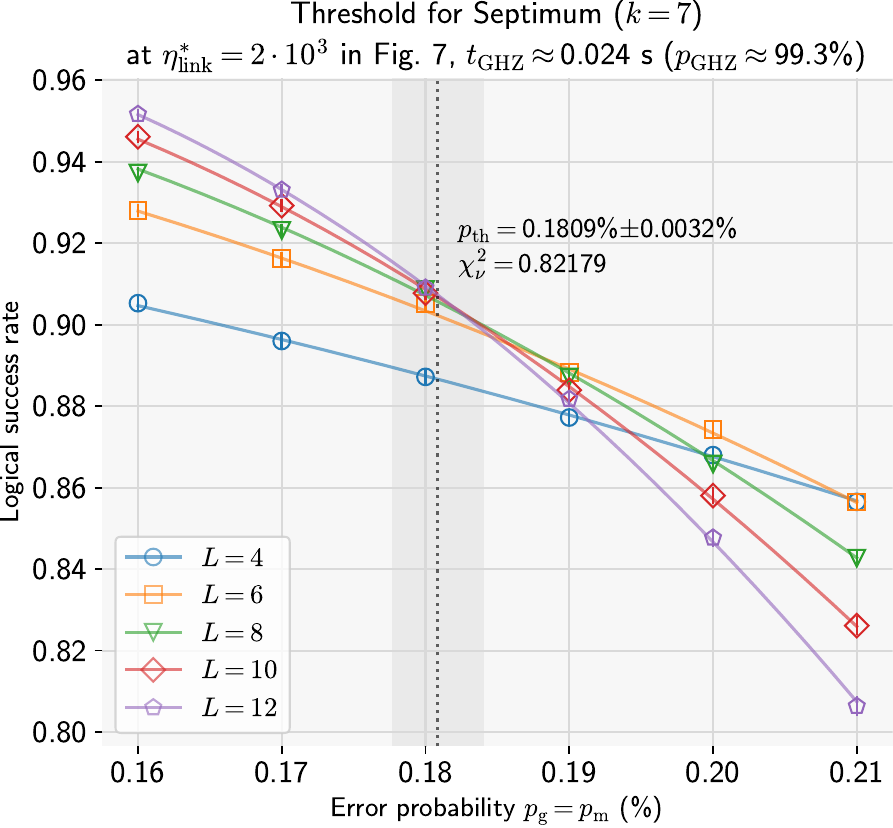}
\caption{Example of a threshold plot for the Septimum protocol, including a 95\% confidence interval calculated with the method described in App.~\ref{subsec:uncertainty_fitting_parameter}.
}
\label{fig:threshold_plot_Septimum_Set3e}
\end{figure}

\begin{figure*}
\centering
\includegraphics[width=\textwidth]{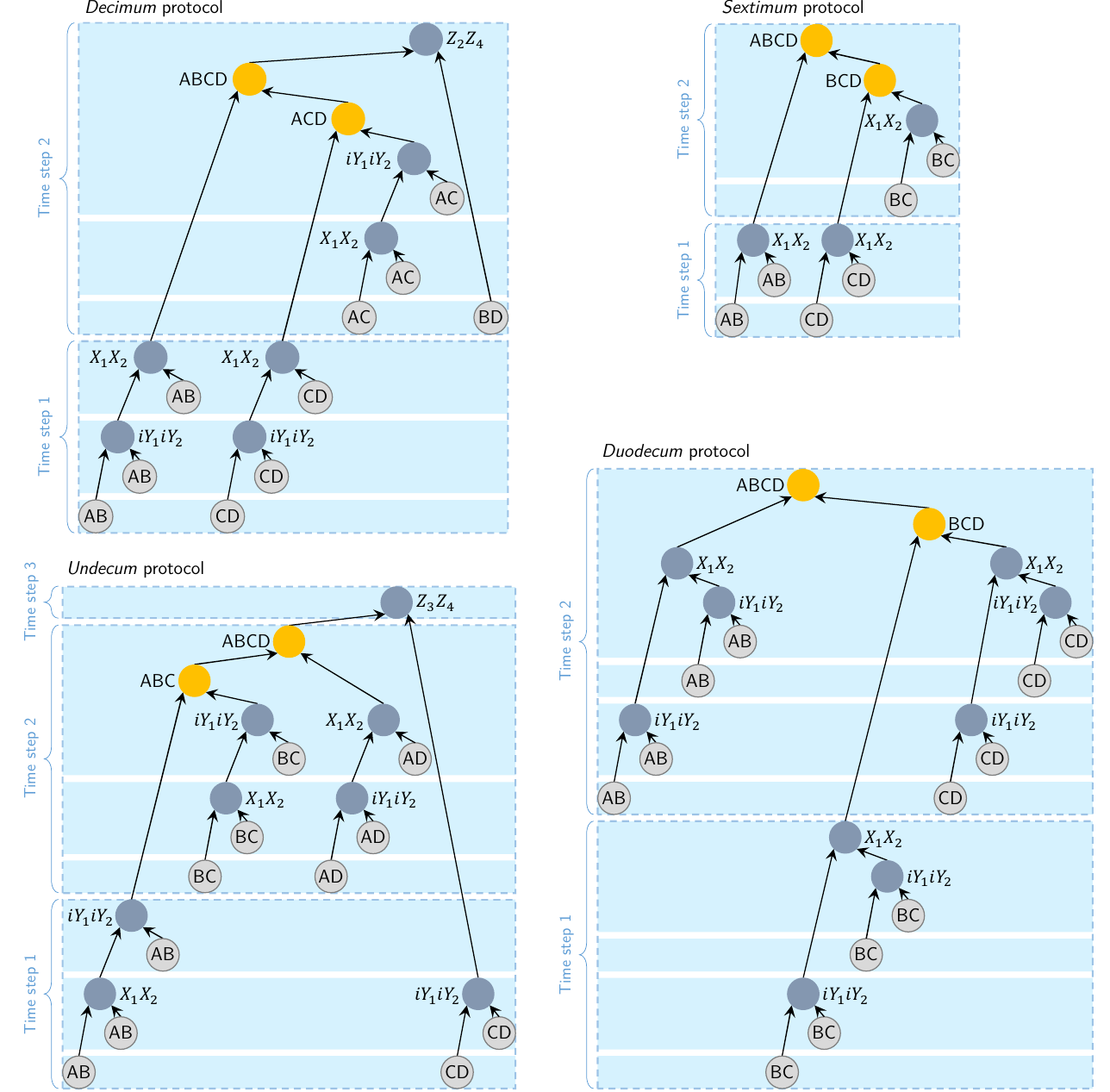}
\caption{Timed binary trees of a selection of best-performing protocols in the Bell pair fidelity and the link efficiency sensitivity studies of Sec.~\ref{subsec:results_near_term}. All four protocols are found with the dynamic program of Sec.~\ref{subsec:dynamic_program_to_generate_GHZ_protocols} for the simulation model and parameters used in Sec.~\ref{subsec:results_near_term}. Clarification on the notation can be found in Fig.~\ref{fig:protocol}. More information on these protocols can be found in Table~\ref{table:best_performing_protocols}. The associated protocol recipes used with these protocols can be found in the repository of Ref.~\onlinecite{deboneDataSoftwareUnderlying2024}.}
\label{fig:best_performing_protocols}
\end{figure*}

The parameter $Q$ from Eq.~\eqref{eq:q_parameter} can also be rewritten as a matrix product:
\begin{equation}
\begin{split}
Q&=\big(\boldsymbol{\epsilon}^{(t+1)}\big)^T\boldsymbol{\Sigma}^{-1}\boldsymbol{\epsilon}^{(t+1)}\\
&=\Big(\boldsymbol{\epsilon}^{(t)}-\boldsymbol{J}^{(t)}\boldsymbol{\Delta\beta}^{(t+1)}\Big)^T\boldsymbol{\Sigma}^{-1}\Big(\boldsymbol{\epsilon}^{(t)}-\boldsymbol{J}^{(t)}\boldsymbol{\Delta\beta}^{(t+1)}\Big).
\end{split}
\end{equation}
Here, $\boldsymbol{\Sigma}$ is a diagonal matrix containing the variances of the observed values $r_i$ as $\boldsymbol{\Sigma}\equiv\text{diag}(\sigma_1^2,\sigma_2^2,\dots,\sigma_{n_\text{C}}^2)$. More generally, one can use the full covariance matrix for $\boldsymbol{\Sigma}$ if this information is available. 

To minimize $Q$ with respect to $\boldsymbol{\Delta\beta}^{(t+1)}$, one sets $\partial Q/\partial\boldsymbol{\Delta\beta}^{(t+1)}$ to zero. This results in an expression that can be solved for $\boldsymbol{\Delta\beta}^{(t+1)}$:~\cite{constalesChapterExperimentalData2017}
\begin{equation}\label{eq:expression_new_beta}
\boldsymbol{\Delta\beta}^{(t+1)}=\Big(\big(\boldsymbol{J}^{(t)}\big)^T\boldsymbol{\Sigma}^{-1}\boldsymbol{J}^{(t)}\Big)^{-1}\big(\boldsymbol{J}^{(t)}\big)^T\boldsymbol{\Sigma}^{-1}\boldsymbol{\epsilon}^{(t)}.
\end{equation}
In obtaining Eq.~\eqref{eq:expression_new_beta}, one has to assume that $\boldsymbol{J}^{(t)}$ does not depend on $\boldsymbol{\Delta\beta}^{(t+1)}$---it is exactly the incorrectness of this assumption for non-linear regression models that makes that the least-squares fitting parameters have to be found iteratively.

\subsection{Uncertainty in fitting parameter values}\label{subsec:uncertainty_fitting_parameter}
The idea of the fitting procedure is that, after sufficient iterations $t\gg1$, $\boldsymbol{\beta}^{(t+1)}$ describes the final (converged) parameter values $\hat{\boldsymbol{\beta}}$. Of course, in theory, in the limit $n_\text{C}\rightarrow\infty$, the fitting parameter values $\hat{\boldsymbol{\beta}}$ would converge to the \emph{true} values, which we indicate with $\boldsymbol{\beta}$. 
One can use another Taylor expansion to express $\hat{\boldsymbol{\beta}}$ in terms of the true set of values $\boldsymbol{\beta}$:
\begin{equation}\label{eq:expression_beta_hat_in_terms_of_beta}
\hat{\boldsymbol{\beta}}\approx\boldsymbol{\beta}+\Big(\hat{\boldsymbol{J}}^T\boldsymbol{\Sigma}^{-1}\hat{\boldsymbol{J}}\Big)^{-1}\hat{\boldsymbol{J}}^T\boldsymbol{\Sigma}^{-1}\boldsymbol{\epsilon}.
\end{equation}
Here, $\boldsymbol{\epsilon}$ describes the residuals $\epsilon_i=r_i-f(\beta,X_i)$ with respect to the true values of the fitting parameter. Furthermore, $\hat{\boldsymbol{J}}$ indicates the Jacobian evaluated with the calculated values $\hat{\boldsymbol{\beta}}$. 
To get Eq.~\eqref{eq:expression_beta_hat_in_terms_of_beta}, one has to assume that $\hat{\boldsymbol{\beta}}\equiv\boldsymbol{\beta}^{(t+1)}=\boldsymbol{\beta}^{(t)}$ holds---\textit{i.e.}, the system of equations has fully converged.

Strictly speaking, in the limit $n_\text{C}\rightarrow\infty$, we have $\hat{\boldsymbol{\beta}}=\boldsymbol{\beta}$ and $\hat{\boldsymbol{\beta}}$ does not have a distribution, since $\boldsymbol{\beta}$ contains constant values. For finite $n_\text{C}$, however, we can argue that $\hat{\boldsymbol{\beta}}$ \emph{does} have a distribution, and we use the last term of Eq.~\eqref{eq:expression_beta_hat_in_terms_of_beta} to estimate the uncertainty in the fitting parameters $\hat{\boldsymbol{\beta}}$. This estimation again involves the assumption that $\hat{\boldsymbol{J}}$ is a constant matrix, and does not depend on the fitting parameters. 

Under this assumption for $\hat{\boldsymbol{J}}$, we can make use of the fact that, for a general constant matrix $\boldsymbol{A}$, the variance of $\boldsymbol{AY}$ is given by $\text{Var}(\boldsymbol{AY})=\boldsymbol{A}\,\text{Var}(\boldsymbol{Y})\boldsymbol{A}^T$. Together with the assumption that $\text{Var}(\boldsymbol{\epsilon})$ can be estimated by $\text{Var}(\boldsymbol{\epsilon})=\boldsymbol{\Sigma}$, the covariance matrix of $\hat{\boldsymbol{\beta}}$ can be expressed as~\cite{ruckstuhlIntroductionNonlinearRegression2010, constalesChapterExperimentalData2017}
\begin{equation}\label{eq:variance_fitting_parameter_values}
\text{Var}\big(\hat{\boldsymbol{\beta}}\big)\approx\Big(\hat{\boldsymbol{J}}^T\boldsymbol{\Sigma}^{-1}\hat{\boldsymbol{J}}\Big)^{-1}.
\end{equation}

The fact that we have a good idea of the uncertainties $\{\sigma_i\}_i$ of the observed $\{r_i\}_i$ means that we are able to estimate the quality of the obtained fit. 
The quality of the fit can be evaluated with the \emph{reduced chi-squared} metric, which is defined as
\begin{equation}
\chi_\nu^2=\frac{Q}{\nu}=\frac{1}{\nu}\sum_{i=1}^{n_\text{C}}\bigg(\frac{r_i-\hat{r}_i}{\sigma_i}\bigg)^2.
\end{equation}
Here, $\nu=n_\text{C}-n_\text{P}$ describes the number of degrees of freedom of the fitting model. A $\chi_\nu^2$ of approximately one corresponds to the variance in the observations matching the variance of the residuals. On the other hand, $\chi_\nu^2<1$ indicates that the uncertainty of the model is too small to describe the data (indicating that the number of fitting parameters might be too large), whereas $\chi_\nu^2>1$ indicates that the model does not describe the data well enough. 

For our fits, we were predominantly interested in the fitting parameter $\hat{p}_\text{th}$ that indicates the threshold value of a certain configuration. We found that the regression model of Eq.~\eqref{eq:fitting_model_equation} worked relatively well in a close range around the true threshold value (see, \textit{e.g.}, Fig.~\ref{fig:threshold_plot_Septimum_Set3e}). If using data over a larger range of $p$ values, we would typically find fits with $\chi_\nu^2>1$. In those situations, we scaled up each $\sigma_i$ with $\chi_\nu$, leading to
\begin{equation}
\text{Var}\big(\hat{\boldsymbol{\beta}}\big)=\chi_\nu^2\Big(\hat{\boldsymbol{J}}^T\boldsymbol{\Sigma}^{-1}\hat{\boldsymbol{J}}\Big)^{-1}, \;\;\text{if}\;\chi_\nu^2>1.
\end{equation}

If one is in possession of $\text{Var}\big(\hat{\boldsymbol{\beta}}\big)$, the standard deviation of the least-squares fitting parameter value $\hat{p}_\text{th}$ can be obtained from the square root of the corresponding diagonal element in $\text{Var}\big(\hat{\boldsymbol{\beta}}\big)$. 
One can then calculate confidence intervals for the fitting parameters by identifying with what factor the standard deviations should be multiplied to ensure the requested level of confidence. For this, we made use of the probability distribution $f_\text{PD}$ of Student's $t$-distribution:
\begin{equation}\label{eq:students_t_distribution}
\begin{split}
f_\text{PD}(t_\text{ci})=\Gamma'(\nu)\bigg(1+\frac{t_\text{ci}^2}{\nu}\bigg)^{-(\nu+1)/2},\\
\Gamma'(\nu)=
\begin{dcases}
    \frac{(\nu-1)(\nu-3)\cdots5\cdot3}{2\sqrt{\nu}(\nu-2)(\nu-4)\cdots4\cdot2},	& \text{if } \nu>1 \text{ even},\\
    \frac{(\nu-1)(\nu-3)\cdots4\cdot2}{\pi\sqrt{\nu}(\nu-2)\nu-4)\cdots5\cdot3},              & \text{if } \nu>1 \text{ odd}.
\end{dcases}
\end{split}
\end{equation}
More specifically, confidence intervals can be calculated by finding the $t_\text{ci}$ factor that corresponds to the confidence interval of choice for the distribution of Eq.~\eqref{eq:students_t_distribution}. 
In the plots in this paper, we show 95\% confidence intervals. For large $\nu$ and a confidence interval of $I_\text{ci}=95$\%, we have $t_\text{ci}\approx1.96$. Smaller values of $\nu$ lead to $t_\text{ci}$ values that are slightly bigger. In Fig.~\ref{fig:threshold_plot_Septimum_Set3e}, we see an example of a threshold plot with a 95\% confidence interval in the value found for the threshold fitting parameter.

\section{Selection of best-performing GHZ generation protocols}\label{app:best_performing_protocols}
At the end of Sec.~\ref{subsec:varying_link_efficiency}, we discuss the Septimum protocol (depicted in Fig.~\ref{fig:protocol}): a GHZ generation protocol found with the dynamic program of Sec.~\ref{subsec:dynamic_program_to_generate_GHZ_protocols} that gives rise to the highest thresholds for the simulation parameters used in several segments of Figs.~\ref{fig:bell_state_quality_plot},~\ref{fig:link_efficiency_plot} and~\ref{fig:bell_succ_prob_plot}. In this appendix, we identify four additional GHZ generation protocols that perform the best in multiple segments of the figures in Sec.~\ref{subsec:results_near_term}: the protocols Sextimum, Decimum, Undecum, and Duodecum. We depict the timed binary trees of these four protocols in Fig.~\ref{fig:best_performing_protocols} and provide more information on their performance in Table~\ref{table:best_performing_protocols}.

\begin{table}[H]
\begin{tabularx}{\columnwidth}{|c||c c c c c|}
\multicolumn{1}{c}{} & \multicolumn{1}{c}{$k$} & \multicolumn{1}{c}{$q$} & \multicolumn{1}{c}{\textbf{Fig.~\ref{fig:bell_state_quality_plot}}} & \multicolumn{1}{c}{\textbf{Fig.~\ref{fig:link_efficiency_plot}}} & \multicolumn{1}{c}{\textbf{Fig.~\ref{fig:bell_succ_prob_plot}}} \\ \cline{1-6}
Sextimum & 6 & 3 & & $[4\cdot10^2, 5\cdot10^2]$ & $4.7\cdot10^2$ \\ \cline{2-6}
Septimum & 7 & 3 & $0.96$ & $[5\cdot10^2, 2\cdot10^3]$ & $[6.3\cdot10^2,1.1\cdot10^3]$ \\ \cline{2-6}
Decimum & 10 & 3 & $[0.9, 0.93]$ & $[2\cdot10^3, 2\cdot10^5]$ & $2\cdot10^3$ \\ \cline{2-6}
Undecum & 11 & 3 & $[0.85, 0.9]$ & & \\ \cline{2-6}
Duodecum & 12 & 4 & $[0.78, 0.82]$ & & \\ \cline{1-6}
\end{tabularx}
\caption{Details about the GHZ generation protocols depicted in Figs.~\ref{fig:protocol} and~\ref{fig:best_performing_protocols}. These protocols are found with the dynamic program of Sec.~\ref{subsec:dynamic_program_to_generate_GHZ_protocols}. The numbers $k$ and $q$ denote the minimum number of Bell pairs and the maximum number of qubits per node required to generate the GHZ state, respectively. The last three columns of the table denote locations or ranges on the $x$-axes of Figs.~\ref{fig:bell_state_quality_plot},~\ref{fig:link_efficiency_plot} and~\ref{fig:bell_succ_prob_plot} in which these protocols are either the best-performing protocol or one of the best-performing protocols.}
\label{table:best_performing_protocols}
\end{table}

\end{document}